\documentclass{article} 
\usepackage[preprint]{neurips_2026}

\usepackage{amsmath,amsfonts,bm}

\def\eqref#1{equation~\ref{#1}}

\def\1{\bm{1}}

\DeclareMathAlphabet{\mathsfit}{\encodingdefault}{\sfdefault}{m}{sl}
\SetMathAlphabet{\mathsfit}{bold}{\encodingdefault}{\sfdefault}{bx}{n}

\usepackage{hyperref}
\usepackage{url}
\usepackage{wasysym}

\usepackage{mathtools} 
\usepackage{booktabs} 
\usepackage{tikz} 
\usepackage{amsthm}
\usepackage{amsmath}
\usepackage{amssymb}
\usepackage[capitalize,noabbrev]{cleveref}
\usepackage{xcolor}
\usepackage{pdfpages}
\usepackage{subcaption}

\usepackage{multirow}
\usepackage{tabularx}

\usepackage[
raggedright
]{sidecap}
\sidecaptionvpos{figure}{t}
\usepackage{afterpage}
\definecolor{mplgreen}{HTML}{008000}
\definecolor{mplblue}{HTML}{0343df}
\definecolor{mplred}{HTML}{FF0000}
\definecolor{mplorange}{HTML}{FFA500}
\definecolor{mpltan}{HTML}{D2B48C}
\definecolor{mplbrown}{HTML}{A52A2A}
\definecolor{mplpurple}{HTML}{800080}

\newcommand\sbullet[1][.5]{\mathbin{\vcenter{\hbox{\scalebox{#1}{$\bullet$}}}}}

\newcounter{imagepage}
\newcommand*{\foreachpage}[2]{%
  \begingroup
    \sbox0{\includegraphics{#1}}%
    \xdef\foreachpage@num{\the\pdflastximagepages}%
  \endgroup
  \setcounter{imagepage}{0}%
  \@whilenum\value{imagepage}<\foreachpage@num\do{%
    \stepcounter{imagepage}%
    #2\relax
  }%
}
\makeatother

\theoremstyle{plain}

\theoremstyle{definition}

\theoremstyle{remark}

\usepackage[noend]{algpseudocode}

\usepackage[textsize=tiny]{todonotes}

\usepackage{changes}

\usepackage{natbib}
\usepackage{verbatim}
\usepackage{pgfkeys}
\usepackage{algorithm}

\usepackage[cal=cm,scr=dutchcal]{mathalpha}

\usepackage{nicematrix}
\usepackage{setspace}
\graphicspath{{./figures/}}

\newcommand{\bigzero}{\mbox{\normalfont\Large\bfseries 0}}

\theoremstyle{definition}
\newtheorem{exmp}{Example}[section]
\newtheorem{exmpnested}{Example}[exmp]

\newcommand{\bz}{\mathbf{z}}
\newcommand{\bb}{\mathbf{b}}

\renewcommand{\bb}{\mathbf{b}}
\newcommand{\bA}{\mathbf{A}}
\newcommand{\bM}{\mathbf{M}}

\newcommand{\bW}{\mathbf{W}}
\newcommand{\bSigma}{\boldsymbol{\Sigma}}
\newcommand{\bmu}{\boldsymbol{\mu}}
\newcommand{\bsigma}{\boldsymbol{\sigma}}
\newcommand{\bL}{\mathbf{L}}
\newcommand{\bu}{\mathbf{u}}
\newcommand{\bv}{\mathbf{v}}
\newcommand{\bw}{\mathbf{w}}

\newcommand{\dd}{\mathrm{d}}

\newcommand{\bV}{\mathbf{V}}
\newcommand{\bJ}{\mathbf{J}}
\newcommand{\bI}{\mathbf{I}}
\newcommand{\bd}{\mathbf{d}}

\newcommand{\bx}{\mathbf{x}}
\newcommand{\calD}{\mathcal{D}}
\newcommand{\calJ}{\mathcal{J}}
\newcommand{\bbeta}{\boldsymbol\beta}

\newcommand{\iid}{\overset{\mathrm{iid}}{\sim} }
\newcommand{\btheta}{\boldsymbol\theta}
\newcommand{\bphi}{\boldsymbol\phi}

\title{Perturbative adaptive importance sampling for Bayesian LOO cross-validation}

\author{
  Joshua C.~Chang\thanks{Correspondence: \texttt{josh.chang@nih.gov}} \\
  NIH Clinical Center\\
  Rehabilitation Medicine\\
  Epidemiology and Biostatistics Section
  \And
  Xiangting Li \\
  UCLA Department of\\Computational Medicine\\
  Los Angeles, CA, USA
  \And
  Tianyi Su \\
  UCLA Department of Statistics\\
  Los Angeles, CA 90095, USA
  \And
  Shixin Xu \\
  Data Science Research Center\\
  Duke Kunshan University\\
  Kunshan, Jiangsu, China
  \And
  Hao-Ren Yao \\
  NIH Clinical Center\\
  Rehabilitation Medicine\\
  Epidemiology and Biostatistics Section
  \And
  Julia Porcino \\
  NIH Clinical Center\\
  Rehabilitation Medicine\\
  Epidemiology and Biostatistics Section
  \And
  Carson Chow \\
  NIH NIDDK
}

\begin{document}

\maketitle

\begin{abstract}

	Importance sampling (IS) is an efficient stand-in for model refitting in performing (LOO) cross-validation (CV) on a Bayesian model.
    IS inverts the Bayesian update for a single observation by reweighting posterior samples.
	The so-called importance weights have high variance -- we resolve this issue through adaptation by transformation.
	We observe that removing a single observation perturbs the posterior by $\mathcal{O}(1/n)$, motivating bijective transformations of the form $T(\btheta)=\btheta + h Q(\btheta)$ for $0<h\ll 1.$
	We introduce several such transformations:
	partial moment matching, which generalizes prior work on affine moment-matching with a tunable step size;
	log-likelihood descent, which partially invert the Bayesian update for an observation;
	and gradient flow steps that minimize the KL divergence or IS variance.
	The gradient flow and likelihood descent transformations require Jacobian determinants, which are available via auto-differentiation; we additionally derive closed-form expressions for logistic regression and shallow ReLU networks.
	We tested the methodology on classification using sparse models ($n\ll p$), count regression (Poisson and zero-inflated negative binomial), and survival analysis problems, finding that no single transformation dominates but their combination nearly eliminates the need to refit.

\end{abstract}
\section{Introduction}
In Bayesian workflows, multiple models are often fitted to a given dataset, and a selection procedure is applied to predict which model will be the most consistent with future observations.
Prediction accuracy is most naturally estimated using cross-validation (CV) of which many variants exist -- the simplest and most ubiquitous of these methods is train-test splitting.
However, estimates of out-of-sample model metrics using train-test splitting are noisy~\citep{dietterichApproximateStatisticalTests1998, kohaviStudyCrossvalidationBootstrap1995} unless computationally expensive k-fold cross-validation (i.e., fitting the model multiple times across the entire dataset) is employed~\citep{rodriguezSensitivityAnalysisKFold2010, wongReliableAccuracyEstimates2020}.

Although $n$-fold -- also known as leave one out (LOO) -- CV is the most expensive of $k$-fold estimators, there exist computationally efficient LOO techniques that completely avoid refitting.
For example, the Akaike Information criteria (AIC) and Bayesian variants~\citep{stoneAsymptoticEquivalenceChoice1977,watanabeAsymptoticEquivalenceBayes2010,gelmanUnderstandingPredictiveInformation2014,watanabeWidelyApplicableBayesian2013} are asymptotic approximations of LOO-CV.
For Bayesian models, a more precise way to compute LOO-CV is to use importance sampling~\citep{vehtariPracticalBayesianModel2017,piironenComparisonBayesianPredictive2017}, which works by using the full data posterior measure as a proposal distribution for each data point's LOO posterior measure.
However, in cases where the LOO measure and full measure are very different, importance sampling can fail~\citep{piironenComparisonBayesianPredictive2017}.
To ameliorate this possibility, we introduce perturbative transformations $T(\btheta)=\btheta+hQ(\btheta)$ that bring the proposal distribution closer to each LOO posterior, exploiting the fact that removing one observation is an $\mathcal{O}(1/n)$ perturbation.
We propose several such transformations:
partial moment matching (PMM), which generalizes existing affine transformations with a tunable step size;
log-likelihood (LL) descent, which directly inverts the Bayesian update at $\mathcal{O}(1)$ cost per observation;
and gradient flow steps that minimize the IS variance or KL divergence.
The gradient-based transformations are model-dependent but are easily computable using autodifferentiation.

\section{Preliminaries}

\paragraph{Notation}
Bold lowercase denotes column vectors, bold uppercase matrices; $\bw_{i,}$ and $\bw_{,j}$ are rows and columns of matrix $\bW$.
Training data $\mathcal{D} = \{ \bd_i\}_{i=1}^n = \{ (\bx_i, y_i)\}_{i=1}^n$; $\calD^{(-i)} = \calD\setminus\{ \bd_i \}$.
Posterior expectations $\mathbb{E}_{\btheta | \calD }$ and LOO expectations $\mathbb{E}_{\btheta | \calD^{(-i)}}$.
For a transformation $T:\Omega\to\Omega$ ($\Omega\subset\mathbb{R}^p$), $\bJ_{T}=\nabla T$ is the Jacobian matrix and $\mathcal{J}_{T}=|\bJ_{T}|$ its determinant.
$\nabla\mu$ yields a column vector, $\nabla\nabla\mu$ the Hessian, and $\nabla^2\mu$ the Laplacian.
$|\cdot|$ denotes the determinant, 2-norm, or absolute value depending on context.

\begin{SCfigure}

	\caption{\textbf{Relationships between probability densities.} One wants to sample from $\pi(\btheta|\calD^{(-i)}),$ the LOO distribution for observation $i$, by sampling from the full-data posterior $\pi(\btheta|\calD).$ The transformation $T_i$ on the full-data posterior brings the sampling distribution closer to the target LOO distribution.}
	\label{fig:mappings}
	\includegraphics[width=0.5\linewidth]{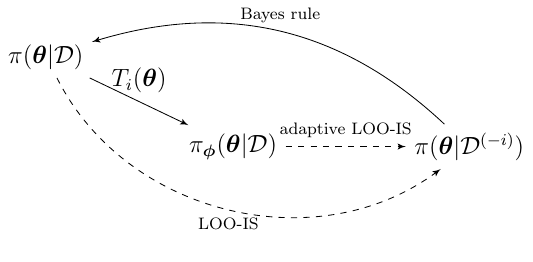}

\end{SCfigure}

\paragraph{Importance sampling for LOO-CV}

Suppose that one has pre-trained a Bayesian  model such that one is able to sample its posterior parameters
$
	\btheta_s \iid \pi(\btheta|\calD).
$
Our objective is to use knowledge of this full-data posterior distribution to estimate how the model would behave if any single point is left out at training.
The full-data posterior and the LOO posterior for observation $i$ are related by the Bayesian update equation, which directly yields the density ratio between the two:
\begin{equation}
	\pi(\btheta | \calD) = \frac{\ell(\btheta|\bd_i) \pi(\btheta | \calD^{(-i)})}{ \int\ell(\btheta|\bd_i) \pi(\btheta | \calD^{(-i)} ) \dd\btheta}
	\quad\implies\quad
	\frac{\pi(\btheta | \calD^{(-i)}) } {\pi(\btheta | \calD)}  = \frac{\mathbb{E}_{\btheta|\calD^{(-i)}}[\ell(\btheta|\bd_i)]}{\ell(\btheta|\bd_i)}\equiv \nu_i(\btheta).
	\label{eq:bayesupdate}
\end{equation}
The implied density ratio $\nu_i(\btheta)$ lets us use the full-data posterior to compute LOO statistics via Monte Carlo~\citep{barbuMonteCarloMethods2020,robertMonteCarloStatistical2013}.
Importance Sampling (IS) is a Monte Carlo method where one computes expectations with respect to a target distribution by taking a weighted average of samples with respect to a given proposal distribution.
For an integrable function $f$,
\begin{align}
	\MoveEqLeft[1] \mathbb{E}_{\btheta|\calD^{(-i)}}\left[f(\btheta)\right] =  \int f(\btheta ) \pi(\btheta | \calD^{(-i)})\dd\btheta  =\int f(\btheta ) \frac{\pi(\btheta | \calD^{(-i)})}{\pi(\btheta | \calD)} \pi(\btheta | \calD)\dd\btheta =\mathbb{E}_{\btheta|\calD} \left[ f(\btheta) \nu_i(\btheta)\right].
	\label{eq:expect}
\end{align}
We approximate Eq.~\ref{eq:expect} by sampling $\btheta_k\iid\pi(\btheta | \calD)$ and computing the self-normalized importance sampling estimator
\begin{equation}
	\mathbb{E}_{\btheta|\calD^{(-i)}}[f(\btheta)] \approx \sum_{k=1}^s \nu_{ik} f(\btheta_k),
	\quad\textrm{where}\quad
	\nu_{ik} = \frac{(\ell(\btheta_k | \bd_i))^{-1}}{\sum_{j=1}^s(\ell(\btheta_j|\bd_i))^{-1}},
	\label{eq:is}
\end{equation}
where the undetermined constant $\mathbb{E}_{\btheta|\calD^{(-i)}}\left[\ell(\btheta|\bd_i)\right]$ cancels in the self-normalized weights $\nu_{ik}$~\citep{gelfandModelDeterminationUsing1992}.

The Bayesian LOO information criterion (LOO-IC), of which the Aikaike Information Criterion (AIC) is an asymptotic approximation~\citep{stoneAsymptoticEquivalenceChoice1977}, can be computed via:
\begin{align}
	{\textrm{LOO-IC}} =-2\sum_{i=1}^n\log\mathbb{E}_{\btheta | \calD^{(-i)}}\left[\ell(\btheta|\bd_i) \right] \approx-2\sum_{i=1}^n\log \sum_{k=1}^s {\nu}_{ik}\ell(\btheta_k|\bd_i).
	\label{eq:looic}
\end{align}

For classification problems, the out-of-sample area under the receiver operator curve or the precision-recall curve is often required.
The LOO AUROC can similarly be computed by propagating LOO estimates of the outcome probabilities
$
	\widehat{\mathscr{p}}_{\textrm{loo},i} =\mathbb{E}_{\btheta | \calD^{(-i)}}[\mathscr{p}_i(\btheta)] \approx \sum_{k=1}^s {\nu}_{ik} \mathscr{p}(\btheta_k, \bx_i).
$

\paragraph{Weight stabilization}
Using the computed posterior $\pi(\btheta|\calD)$ as the proposal distribution often yields slow convergence -- the $1/\ell$ importance weights, being fat tailed, have large or unbounded variance~\citep{peruggiaVariabilityCaseDeletionImportance1997}, making the IS estimate for LOO-CV expectations (Eq.~\ref{eq:is}) noisy.

Two practical model agnostic methods for controlling the tail of importance weights are weight truncation~\citep{ionidesTruncatedImportanceSampling2008} and Pareto smoothing~
\citep{vehtariParetoSmoothedImportance2024,vehtariPracticalBayesianModel2017}.
Pareto smoothing replaces the largest $M$ weights with their corresponding rank-values from a fitted generalized Pareto-distribution~\citep{zhangNewEfficientEstimation2009}.
Pareto smoothed importance sample (PSIS)-based LOO-CV implementations are widely available in software packages such as \texttt{Stan} and \texttt{ArviZ}.
However, PSIS-LOO fails when the tail distribution of importance weights is not well-fit by the Pareto distribution; a general rule of thumb is that the parameter $\hat{k}$ exceeds $0.7$.
In these cases, performing an additional model-specific controlled transformation on the proposal distribution will induce more efficient computations.
	{Later on, as in \citet{paananenImplicitlyAdaptiveImportance2021}, we will use the estimated Pareto shape parameter $\hat{k}$ on post-transformation IS weights in order to evaluate the success of different transformations.
	Effective transformation should be able to reduce the Pareto shape parameter to below the given threshold.}

\paragraph{Adaptive importance sampling (AIS)}

We apply the concept of AIS~\citep{bugalloAdaptiveImportanceSampling2017,cornuetAdaptiveMultipleImportance2011,elviraAdvancesImportanceSampling2022} to transform the posterior distribution to be closer to the LOO-CV distribution $\pi(\btheta|\calD^{(-i)})$ (the relationships between the different distributions are depicted in Fig.~\ref{fig:mappings}).

Consider the bijection $T_i:\mathbb{R}^p\to\mathbb{R}^p$, defined for observation $i,$ and let $\bphi \equiv T_i(\btheta).$
By change of variables,
$
	\pi_{\bphi}(\bphi|\ldots) = \pi\left(T_i^{-1}(\bphi)|\ldots \right)\mathcal{J}_i^{-1}(\bphi),\label{eq:changeofvars}
$
where we denote $\bJ_T = \nabla T$,
$\calJ_{T_i}^{-1}(\bphi) =\left|\bJ_{T_i}^{-1} (\bphi)\right|,
$
and
$
	\mathcal{J}_{T_i}(\btheta) = \left|\bJ_{T_i} (\btheta)\right| =  {1}/{\mathcal{J}_{T_i}^{-1}(\bphi)},
$
%
The expectation in Eq.~\ref{eq:expect} in terms of an integral over $\pi_{\bphi}$ is
\begin{align} 
	\mathbb{E}_{\btheta|\calD^{(-i)}}\left[f(\btheta)\right] & = \int f(\btheta) \nu_i(\btheta)\pi(\btheta |\calD)\dd\btheta   = \int f(\btheta) \nu_i(\btheta) \frac{\pi(\btheta |\calD)}{\pi_{\bphi}(\btheta |\calD)}
	\pi_{\bphi}(\btheta |\calD)\dd\btheta \nonumber                                                                                                                                                                     \\
	                                                         & = \int f(\btheta)\nu_i(\btheta)
	\frac{\pi(\btheta |\calD)\calJ_{T_i}(T_i^{-1}(\btheta))}
	{\pi(T_i^{-1}(\btheta) |\calD)}
	\pi_{\bphi}(\btheta|\calD)\dd\btheta.
	\label{eq:expect2}
\end{align}
Define a Monte Carlo approximation of Eq.~\ref{eq:expect2} using importance sampling, by sampling $\btheta_k\iid \pi(\btheta| \calD)$ so that $\bphi_k =T_i(\btheta_k)\iid \pi_{\bphi}(\bphi| \calD):$
\begin{align}
	\mathbb{E}_{\btheta|\calD^{(-i)}}\left[f(\btheta)\right] \approx \sum_{k=1}^s \frac{\eta_{ik}}{\sum_{j=1}^s \eta_{ij}} f(\bphi_k) \qquad
	\eta_{ik}= \frac{\calJ_{T_i}(\btheta_k)}{\ell(\bphi_k|\bd_i)}\frac{\pi(\bphi_k|\calD)}{\pi(\btheta_k|\calD)}.\label{eq:importance}
\end{align}
By Bayes rule, the posterior likelihood ratio in Eqs.~\ref{eq:expect2}--\ref{eq:importance}  has the exact expression
${\pi(\bphi|\calD)}/{\pi(\btheta|\calD)} = [{\pi(\bphi)}/{\pi(\btheta)}]\prod_j [{\ell(\bphi|\bd_j)}/{\ell(\btheta|\bd_j)}].$
Computing this expression requires iterating over the entire dataset.
There are various methods to avoid this expensive computation, for instance also using Monte Carlo approximation by sampling data points.
For large datasets, one can turn to variational approximations.

\paragraph{Correcting variational posteriors}

For computational expediency, variational methods are often used in place of MCMC for Bayesian inference, obtaining a variational approximation $\hat{\pi}(\btheta|\calD)$ to the true posterior, where $\hat{\pi}$ lies within a given family of probability distributions.
In problems where one expects a substantial discrepancy between the true posterior and $\hat{\pi},$ one may correct for this discrepancy by noting that
\begin{align} 
	\mathbb{E}_{\btheta|\calD^{(-i)}}\left[f(\btheta)\right] & = \int f(\btheta) \nu_i(\btheta)\pi(\btheta |\calD)\dd\btheta    = \int f(\btheta) \nu_i(\btheta) \frac{\pi(\btheta |\calD)}{\hat{\pi}_{\bphi}(\btheta |\calD)}
	\hat{\pi}_{\bphi}(\btheta |\calD)\dd\btheta \nonumber                                                                                                                                                                      \\
	                                                         & = \int f(\btheta)\nu_i(\btheta)
	\frac{\pi(\btheta |\calD)\mathcal{J}_{T_i}(T_i^{-1}(\btheta))}
	{\hat{\pi}(T_i^{-1}(\btheta) |\calD)}
	\hat{\pi}_{\bphi}(\btheta|\calD)\dd\btheta
	\label{eq:expect3}
\end{align}
and using the self-normalized importance sampler
\begin{align}
	\MoveEqLeft\mathbb{E}_{\btheta|\calD^{(-i)}}\left[f(\btheta)\right] \approx \sum_{k=1}^s \frac{\chi_{ik}}{\sum_{j=1}^s \chi_{ij}} f(\bphi_k)  \qquad \chi_{ik}= \frac{\mathcal{J}_i(\btheta_k)}{\hat{\pi}(\btheta_k|\calD)} \pi(\bphi_k) \prod_{j\neq i}\ell(\bphi_k|\bd_j),\label{eq:variationalimportance}
\end{align}
where $\pi(\bphi_k)$ is the prior density at $\bphi_k$,
canceling out the two unknown constants corresponding to  $\pi(\bphi_k|\calD)$ and $\nu_i.$

\section{Methods}

Eq.~\ref{eq:importance} is valid for an arbitrary bijection $T_i:\mathrm{supp}(\pi(\btheta))\to\mathrm{supp}(\pi(\btheta)).$
The objective of using transformations is to shift the proposal distribution closer to the targeted LOO distribution for each observation -- to invert the  update version of Bayes' rule.
Returning to Eq.~\ref{eq:bayesupdate}, removing a single observation from the posterior is a small perturbation: since $\log\pi(\btheta|\calD) = \log\pi(\btheta|\calD^{(-i)}) + \log\ell(\btheta|\bd_i) + \textrm{const},$
the KL divergence between the full and LOO posteriors satisfies
$\mathrm{D}_{\mathrm{KL}}(\pi(\btheta|\calD^{(-i)})\Vert\pi(\btheta|\calD)) = \log\mathbb{E}_{\btheta|\calD^{(-i)}}[\ell(\btheta|\bd_i)] - \mathbb{E}_{\btheta|\calD^{(-i)}}[\log\ell(\btheta|\bd_i)],$
which is $\mathcal{O}(1/n)$ for well-specified models where each observation contributes $\mathcal{O}(1/n)$ to the posterior.
The $\mathcal{O}(1/n)$ scaling motivates the use of perturbative transformations
$
	T_i(\btheta) = \btheta + hQ_i(\btheta),
$
for a small step size $h>0$ and vector field $Q_i.$
We introduce several such transformations, each nudging the full-data posterior samples toward the LOO-CV distribution via a different mechanism.

\paragraph{Partial moment matching}
\label{sec:momentmethod}
The three transformation methods presented in \citep{paananenImplicitlyAdaptiveImportance2021} match the first two moments  of the proposal distribution and the target distribution, independently for each scalar component of each model parameter.
We generalize those transformations subject to a tunable scalar constant $\bar{h},$
\begin{align}
	T_{\textrm{PMM}1}(\btheta) & = \btheta  + \bar{h}(\bar\btheta_w - \bar\btheta)   & T_{\textrm{PMM}2}(\btheta) & =  \btheta  +  \bar{h}\left(\mathbf{v}_w^{1/2}\circ \bv^{-1/2}\circ(\btheta-\bar\btheta)  + \bar\btheta_w -\btheta \right) \nonumber \\
	\bar\btheta                & = \frac{1}{s}\sum_{k=1}^s\btheta_k                  & \bv                        & = \frac{1}{s}\sum_{k=1}^s (\btheta_k - \bar\btheta)\circ(\btheta_k-\bar\btheta) \nonumber                                            \\
	\bar\btheta_w              & =  \frac{\sum_{k=1}^s \nu_k\btheta_k}{\sum_k \nu_k} & \bv_w                      & = \frac{\sum_{k=1}^s \nu_k (\btheta_k - \bar\btheta)\circ(\btheta_k-\bar\btheta)}{\sum_{k=1}^s \nu_k}
\end{align}
where setting $\bar{h}=1$ recovers the original transformations MM1/MM2 respectively.
We additionally introduce a full-rank generalization PMM3 that matches the complete covariance structure (not just marginal variances), parameterized by the same tunable $\bar{h}$; the full derivation appears in Supplement~\ref{sec:pmm3}.
Setting $\bar{h}=1$ recovers the MM3 transformation of \citet{paananenImplicitlyAdaptiveImportance2021}.

\paragraph{Log-likelihood descent}
\label{sec:lldescent}
The GRAMIS algorithm~\citep{elviraGradientAdaptivePopulation2015,elviraAdvancesImportanceSampling2022,elviraGradientbasedAdaptiveImportance2022} iteratively updates the means and covariances of a population of Gaussian proposals via gradient descent on the likelihood function, constructing a Gaussian mixture approximation to the LOO posterior that serves as a proposal density.

We take a similar approach but perform the descent directly in sample space.
During model training, each observation $\bd_i$ pulls the model parameters in the direction $\nabla_{\btheta}\log \ell(\btheta | \bd_i).$
The LL transformation steps against this direction,
\begin{equation}
	T_i^{\textrm{LL}}(\btheta) = \btheta - h\nabla_{\btheta}\log \ell(\btheta | \bd_i), \label{eq:eulerstepLL}
\end{equation}
shifting posterior samples toward $\pi(\btheta|\calD^{(-i)})$ without requiring an explicit posterior density evaluation.
Because the gradient $\nabla\log\ell$ varies with $\btheta$, the transformation is nonlinear and can preserve non-Gaussian structure (skewness, heavy tails) that affine methods cannot capture.

\paragraph{Gradient flow transformations}

We consider choosing $T_i$ to minimize either (a) the variance of the IS estimator for a target function $f$, or (b) the KL divergence $\mathrm{D}_{\mathrm{KL}}\left( \pi({\btheta|\calD^{(-i)}}) \Vert \pi_{\bphi}(\btheta|\calD)\right)$.

The IS variance objective (derived in Supplement~\ref{sec:variance}) yields the variance-reducing transformation
\begin{align}
	T^{\textrm{Var}}_i(\btheta) =\btheta + hQ_i^{\textrm{Var}}(\btheta)
	\qquad
	Q_i^{\textrm{Var}}(\btheta) =  \pi(\btheta|\calD)\frac{f(\btheta)}{\ell(\btheta|\bd_i)}\nabla\left( \frac{f(\btheta)}{\ell(\btheta|\bd_i)}\right)
	\label{eq:eulerstepvar0}
\end{align}
where $f(\btheta)$ is a target function matching the estimation objective.
The choice of $f$ requires care: if $f/\ell$ is constant (or nearly so), the gradient $\nabla(f/\ell)$ vanishes and $Q_i^{\textrm{Var}}=0$.
For instance, in sigmoidal models with the natural choice $f=\mathscr{p}_i(\btheta)$, the ratio $f/\ell = 1$ when $y_i=1$, rendering the transformation ineffective for half the observations.
As derived in Supplement~\ref{sec:variance}, the symmetric form $f_i(\btheta) = \mathscr{p}_i(\btheta)^{1-y_i}(1-\mathscr{p}_i(\btheta))^{y_i}$ avoids this cancellation.
For count and survival models we use $f_i(\btheta) = P(Y_i \leq t_0 | \btheta)$ for a chosen quantile $t_0$; more-general formulae are in the Supplement.

Minimizing the KL divergence is equivalent to minimizing the cross-entropy
$
	H\left( \pi({\btheta|\calD^{(-i)}}), \pi_{\bphi}(\btheta|\calD)\right) =   -\int \nu_i(\bphi)\pi(\bphi|\calD) \log [{\pi(T_i^{-1}(\bphi)|\calD)}/{\mathcal{J}_{T_i}(T_i^{-1}(\bphi))}]  \dd\bphi
$
with respect to $T_i.$
The Euler-Lagrange equation (derived in Supplement~\ref{sec:KLdivergence}) is implicit in $T_i,$ but $T_i$ is a stable fixed point of the gradient flow
${\partial T_i}/{\partial t} = - {\delta H}/{\delta T_i}.$
Taking a single forward Euler step from $T_i(\btheta)=\btheta$ with step size $h[\mathbb{E}_{\btheta|\calD^{(-i)}}[\ell(\btheta|\bd_i)]]^{-1}$ yields
\begin{align}
	T^{\textrm{KL}}_i(\btheta) = \btheta +h\underbrace{\pi({\btheta}|\calD)\nabla\left(\frac{1}{\ell(\btheta|\bd_i)} \right)}_{Q_i^{\textrm{KL}}}.
	\label{eq:eulerstepKL}
\end{align}

\paragraph{Resolving the posterior density} Both the KL (Eq.~\ref{eq:eulerstepKL})  and variance (Eq.~\ref{eq:eulerstepvar0}) descent transformations take steps proportional to the posterior density $\pi(\btheta|\calD).$
If a variational approximation for $\pi({\btheta}|\calD)$ is available, using it in  Eqs.~\ref{eq:eulerstepKL} and~\ref{eq:eulerstepvar0}  as a stand-in for the posterior density helps simplify the computation of the transformations and their Jacobians, particularly when using mean-field or low-order Automatic Differentiation Variational Inference (ADVI)~\citep{kucukelbirAutomaticDifferentiationVariational2017,bleiVariationalInferenceReview2017}.

In the absence of variational approximation,
one may evaluate the posterior densities exactly using the Bayes rule, absorbing the unknown normalization constant $\mathcal{Z}$ into the step size $h.$
The obvious downside of using these exact transformations is the need to iterate over the entire dataset in order to evaluate the posterior density, which must be done for each parameter sample, for each data point.
Note that the LL transformation (Eq.~\ref{eq:eulerstepLL}) sidesteps this issue entirely, as it depends only on the gradient of a single observation's log-likelihood.

For evaluating the Jacobian determinants, one appeals to Bayes rule to find that
$\nabla\log[\mathcal{Z}\pi(\btheta|\calD)] = \nabla\log\pi(\btheta) + \sum_i \nabla\log\ell(\btheta|\bd_i),$
where $\mathcal{Z}$ is absorbed into $h$.

\paragraph{Jacobian determinant approximation}

For either single-step transformations, one may approximate $\left|\bJ_{T_i}\right|$ by noting that
\begin{align}
	\calJ_{T_i}(\btheta) = \left| 1 + h\nabla\cdot Q_i(\btheta)   \right| + \mathcal{O}(h^2)  \label{eq:Japprox}
\end{align}
and truncating to $\mathcal{O}(h),$ sidestepping the computation of Hessian matrices and their spectra.
The exact determinant is $\prod_k(1+h\lambda_k)$ where $\lambda_k$ are eigenvalues of $\nabla Q_i$; the leading error term in log-space is
$\log\calJ^{\textrm{exact}} - \log\calJ^{\textrm{approx}} = {h^2}[(\nabla\cdot Q_i)^2 - \textrm{tr}((\nabla Q_i)^2)]/{2} + \mathcal{O}(h^3),$
which enters the importance weights (Eq.~\ref{eq:importance}) as a multiplicative $(1+\epsilon_k)$ perturbation with $\epsilon_k = \mathcal{O}(\bar{h}^2)$.
Because the self-normalized IS estimator divides by $\sum_k\eta_{ik}$, a constant multiplicative error cancels exactly; the residual bias is proportional to $\mathrm{Cov}_\eta[\epsilon_k, f(\bphi_k)]$, the covariance between the per-sample error and the target function under the importance-weighted distribution (Supplement~\ref{sec:jacobianpropagation}).
For the LL transformation, $Q_i$ does not depend on $\pi(\btheta|\calD)$, so $\epsilon_k$ is approximately constant across samples and the self-normalization cancels the Jacobian error exactly.
For the Var/KL transformations, $\epsilon_k$ varies systematically across samples through $\pi(\btheta|\calD)$ -- modal samples incur larger errors than tail samples -- and the effective bias scales as $\bar{h}^2 k^2/n$ where $k$ is the effective sparsity (Supplement~\ref{sec:jacobianerror}).
For large problems, computing the full spectra of $\nabla Q_i$ is impractical.

\paragraph{Step size selection}\label{sec:stepsize}
For all parameter samples at each individual observation $i,$ we use
$
	h_i = \bar{h}\min_{s, \alpha}\left\{ \left| {\sqrt{\Sigma_{\alpha,\alpha}}}/{Q_i(\btheta_s)_{\alpha} }\right|\right\} \label{eq:hi}
$
where $\bar{h}>0$ and $\sqrt{\Sigma_{\alpha,\alpha}}$ is the marginal posterior standard deviation of the $\alpha$-th component of $\btheta.$
The rule ensures that the transformation takes a step of at most $\bar{h}$ posterior standard deviations in any parameter component.
Since the posterior concentrates at rate $\mathcal{O}(1/\sqrt{n})$, and removing one observation perturbs the posterior by $\mathcal{O}(1/n)$ in parameter space, the natural step size is $\bar{h}\sim 1/\sqrt{n}$ posterior standard deviations.
An influence function analysis (Supplement~\ref{sec:jacobianerror}) confirms this: the LOO perturbation in posterior standard deviation units is $|\Delta\theta^*_\alpha|/\sigma_\alpha \sim 1/\sqrt{n}$ for typical observations, with high-leverage observations exceeding this bound -- precisely those requiring adaptation.
The same scaling emerges from minimizing a second-order expansion of the cross-entropy along the gradient flow direction (Supplement~\ref{sec:jacobianerror}).
However, the Jacobian approximation error of $\mathcal{O}(\bar{h}^2 k^2/n)$ permits $\bar{h}\lesssim\sqrt{n}/k$ before the error exceeds $\mathcal{O}(0.1)$.
For sparse models ($k\ll\sqrt{n}$), step sizes well above $1/\sqrt{n}$ are safe; the upper bound $\bar{h}\geq 1$ exceeds the perturbative regime, while $\bar{h}\ll 1/\sqrt{s}$ produces shifts below the Monte Carlo noise floor.
For the LL and PMM transformations, whose vector fields $Q_i$ do not depend on $\pi(\btheta|\calD)$, every sample receives the same normalized step.
For the KL and Var transformations, $Q_i$ includes a factor of $\pi(\btheta|\calD)$, so samples near the posterior mode receive larger steps while tail samples are implicitly regularized.
In practice we scan $\bar{h}=2^{-r}$ for $r\in\{1, 2,\ldots, 8\}$, covering $\bar{h}$ from $1/2$ down to $\approx 1/256$; this range includes step sizes near $1/\sqrt{n}$ for all datasets considered ($n\in\{54, 262, 686\}$) while staying within the perturbative regime where the first-order Jacobian approximation is accurate.
All step sizes can be evaluated in parallel using vectorized operations.
Although searching over multiple step sizes and transformations involves evaluating many candidates per observation, the $\hat{k}$ diagnostic is a property of the importance weight distribution rather than a statistical test -- a transformation that achieves $\hat{k}<0.7$ produces an IS estimator with finite variance regardless of how many other candidates were tried.

\paragraph{Overview} We have provided six transformations, each aimed at stabilizing a LOO importance sampler by bringing the proposal distribution closer to the LOO target in a different sense. The PMM1/PMM2/PMM3 affine transformations shift the moments of the posterior samples closer to those of the target distribution. The LL descent transformation steps up the log-likelihood gradient. The KL and Var descent transformations take one step along their corresponding gradient flow equations.
Of these transformations, LL is the most computationally efficient, as it avoids the posterior density evaluation.

Generally, many observations are amenable to direct importance sampling with $1/\ell$ weights (Eq.~\ref{eq:expect}) in combination with Pareto smoothing (tail weight distribution shape parameter $\hat{k}<0.7$).
One needs only transform the sampling distribution when $\hat{k}$ exceeds this threshold~\citep{vehtariParetoSmoothedImportance2024}.
For each observation $i$, one iterates through the available transformations, applying each to the posterior samples $\btheta_1,\ldots,\btheta_s\iid\pi(\btheta|\calD)$, recomputing the transformed weights $\eta_{ik}$ (Eq.~\ref{eq:importance}) and $\hat{k}$, and stopping as soon as any transformation achieves $\hat{k}\leq 0.7.$
If \emph{any} transformation succeeds for a given observation, adaptation is successful and one avoids refitting the entire model -- at potentially large time and energy cost savings.

\section{Examples}

The gradient flow and likelihood descent transformations are model-specific, requiring Jacobian terms that are amenable to auto-differentiation.
To put the method into concrete terms, we derive the required Jacobian terms for a ubiquitous category of classification models.
\begin{exmp}[Sigmoidal classification models]
	\label{exmp:sigmoidal_models}

	In these models a vector of covariates $\bx\in\mathbb{R}^p$ is used to estimate the probability of an outcome labeled by $y\in\{0,1\}$
	with likelihood function $\ell$:
	\begin{align}
		y_i | \btheta, \bx_i \ {\sim}
		\ \textrm{Bernoulli}\left(\mathscr{p}_i(  \btheta) \right) \qquad
		\ell(\btheta |  y_i, \bx_i) =  \mathscr{p}_i(  \btheta)^{y_i}   (1-\mathscr{p}_i(  \btheta)  )^{1-y_i},
	\end{align}
	and where $\mathscr{p}_i(\btheta) \equiv \mathscr{p}(\btheta, \bx_i)$ is the predicted outcome probability for observation $i$, with a sigmoidal parameterization
	$
		\mathscr{p}_i(\btheta) = p(\btheta, \bx_i) = \sigma(\mu_i(\btheta))
	$
	where $\sigma(\mu)=1/(1+e^{-\mu})$ is the sigmoid function and we denote $\mu_i(\btheta) \equiv \mu(\btheta,\bx_i)$ for some mean function $\mu$.
	For these models, the vector fields $Q_i$ specialize to
	\begin{align}
		Q_i^{\textrm{KL}}(\btheta) &= (-1)^{y_i}{\pi(\btheta|\calD)}  e^{\mu_i(\btheta)(1-2y_i)}  \nabla\mu_i \nonumber\\
		Q_i^{\textrm{Var}}(\btheta) &= (-1)^{y_i}\pi(\btheta|\calD) e^{2\mu_i(\btheta)(1-2y_i)} \nabla\mu_i \label{eq:Qsigmoid}\\
		Q_i^{\textrm{LL}}(\btheta) &= -\left[y_i\big(1-\sigma(\mu_i(\btheta))\big) - (1-y_i)\sigma(\mu_i(\btheta))\right]\nabla\mu_i \nonumber
	\end{align}
	and their Jacobians (derived in Supplement~\ref{sec:sigmoidjacobian}) depend on the Hessian $\nabla\nabla\mu_i$ and the posterior gradient $\nabla\log\pi(\btheta|\calD)$.
	Here we will consider two popular subfamilies of sigmoidal models.

	\begin{exmpnested}[Logistic Regression (LR)]

		LR is a sigmoidal model where $\mu_i(\btheta)=\bx_i^\intercal{\bbeta}$, so $\nabla_{\bbeta}\mu_i = \bx_i$, and $\nabla\nabla\mu = \mathbf{0}$.
		Because the Hessian of $\mu$ vanishes, the Jacobian of $Q_i$ is rank-one, admitting exact determinants (Supplement~\ref{sec:lrjacobian}).

	\end{exmpnested}

	We also derive explicit expressions for shallow Bayesian ReLU neural networks (Supplement~\ref{sec:reluexample}).
	Their Hessians of $\mu$ are block-sparse -- in particular $\nabla\nabla_{W_k}\mu = \mathbf{0}$ for all layers -- so the first-order Jacobian approximation (Eq.~\ref{eq:Japprox}) can ignore the model Hessian entirely.
	For single hidden layers, the Hessian has exactly $2d$ non-zero eigenvalues (where $d$ is the hidden dimension), admitting exact Jacobian determinants via rank-one updates.

\end{exmp}

\section{Experiments}

Implementations of our method in both Python/JAX~\citep{jax2018github} and \texttt{r/brms}~\citep{burknerBrmsPackageBayesian2017} are available at \texttt{github:mederrata/bayesianquilts}.
Jupyter notebooks for producing the results in the text are included in the Supplement.
As baselines for comparison, we evaluated the original MM1/MM2/MM3 affine transformation methods of \citet{paananenImplicitlyAdaptiveImportance2021}, the GRAMIS algorithm~\citep{elviraGradientAdaptivePopulation2015,elviraAdvancesImportanceSampling2022,elviraGradientbasedAdaptiveImportance2022}  (implementation details in Supplement~\ref{sec:gramis_implementation}), and the mixture importance sampling (MixIS) method of \citet{silvaRobustLeaveoneoutCrossvalidation2024}.
\citet{paananenImplicitlyAdaptiveImportance2021} used a split-sampling scheme; for the most direct comparison between transformations, we evaluated MM1/MM2/MM3 without split sampling.

\paragraph{Datasets and models}
For demonstration, we used models trained on three publicly available datasets: an ovarian cancer micro-array dataset~\cite{hernandez-lobatoExpectationPropagationMicroarray2010,schummerComparativeHybridizationArray1999} consisting of $n=54$ observations of $p=1056 + 1$ predictors, a pest treatment dataset for estimating the effect of a treatment on the number of detected roaches~\citep{paananenImplicitlyAdaptiveImportance2021}, and the GBSG2 breast cancer survival dataset~\citep{schumacherRandomizedClinicalTrial1994} consisting of $n=686$ patients with right-censored survival times.
\citet{paananenImplicitlyAdaptiveImportance2021} used the first two datasets to test their moment-matching adaptive importance sampler; we reproduced their models -- logistic regression (ovarian) and Poisson regression (roach)~\citep{piironenHyperpriorChoiceGlobal2017,piironenSparsityInformationRegularization2017,carvalhoHandlingSparsityHorseshoe2009}.
We also fitted additional models: shallow ReLUnet (ovarian) and zero-inflated negative binomial regression (roach).
For the GBSG2 dataset, we fitted two piecewise exponential survival (PES) models: a neural model with two hidden layers of 16 units ( NPES), and a quilted model~\citep{changInterpretableNotJust2024,xiaInterpretableNotJust2023a} with additive decomposition over five categorical covariates (108 groups $\times$ 5 intervals) with horseshoe shrinkage priors (QPES).
Both survival models use breakpoints at quintiles of uncensored event times.
We purposely over-parameterized the models relative to the dataset size, producing ill-conditioned posteriors where a meaningful fraction of observations require adaptation -- providing a stress test for the methods.

\paragraph{Adaptation}
\begin{table}[h!]
	\centering
	\small
	\setlength{\tabcolsep}{3pt} 
	\begin{tabular*}{\textwidth}{@{\extracolsep{\fill}} ll cc cc cc}
		\toprule
		\multirow{2}{*}{ } & \multirow{2}{*}{\textbf{Method}} & \multicolumn{2}{c}{\textbf{Ovarian}  (n=54)} & \multicolumn{2}{c}{\textbf{Roaches} (n=262)} & \multicolumn{2}{c}{\textbf{GBSG2} (n=686)}\\
		\cmidrule(lr){3-4} \cmidrule(lr){5-6} \cmidrule(lr){7-8}
		& & LR & NN & PR & NBR & NPES & QPES\\
		\midrule
		Base & No Adaptation & $33.8\pm3.1$ & $15.9\pm2.6$ & $49.8\pm5.0$ & $84.8\pm7.6$ & $41.1\pm23.3$ & $4.9\pm1.9$\\
		\midrule
		\multirow{7}{*}{Ours}
		& PMM1 & $21.6\pm3.5$ & $5.9\pm2.2$ & $41.1\pm3.6$ & $2.1\pm1.0$ & $0.8\pm1.0$ & $3.9\pm1.4$\\
		& PMM2 & $33.2\pm3.3$ & $6.3\pm1.8$ & $41.6\pm3.3$ & $24.5\pm10.3$ & $1.5\pm1.6$ & $3.4\pm1.2$\\
		& PMM3 & $30.8\pm3.7$ & $8.2\pm2.4$ & $37.9\pm3.1$ & $32.1\pm8.4$ & $2.4\pm1.4$ & $2.5\pm1.1$\\
		& LL   & $5.7\pm2.0$ & $7.8\pm1.6$ & $33.6\pm2.6$ & $8.6\pm1.9$ & $16.6\pm12.3$ & $4.3\pm1.3$\\
		& Var  & $33.8\pm3.1$ & $10.1\pm1.7$ & $49.8\pm5.0$ & $84.8\pm7.6$ & $41.1\pm23.3$ & $4.9\pm1.9$\\
		& KL   & $27.2\pm2.3$ & $7.8\pm2.8$ & $49.7\pm5.0$ & $83.5\pm7.3$ & $30.0\pm18.5$ & $4.8\pm2.0$\\
		\cmidrule(lr){2-8}
		& \textbf{Combined} & $\mathbf{0.6\pm0.7}$ & $\mathbf{1.7\pm2.7}$ & $\mathbf{25.4\pm2.2}$ & $\mathbf{1.8\pm0.8}$ & $\mathbf{0.2\pm0.4}$ & $\mathbf{2.3\pm1.0}$\\
		\midrule
		\multirow{6}{*}{Prior}
		& MM1 & $30.2\pm2.5$ & $8.2\pm2.1$ & $48.6\pm5.5$ & $65.9\pm9.1$ & $5.5\pm7.1$ & $2.6\pm0.9$\\
		& MM2 & $26.1\pm3.3$ & $8.5\pm1.5$ & $49.6\pm4.9$ & $61.8\pm10.6$ & $2.0\pm3.0$ & $2.4\pm1.1$\\
		& MM3 & $6.8\pm2.4$ & $12.1\pm2.7$ & $49.8\pm5.0$ & $70.1\pm10.3$ & $2.0\pm2.7$ & $4.4\pm2.2$\\
		& GRAMIS & $33.8\pm3.1$ & $15.9\pm2.6$ & $39.0\pm1.8$ & $79.7\pm7.1$ & $41.1\pm23.3$ & $4.9\pm1.9$\\
		& MixIS & $21.9\pm2.6$ & $11.4\pm2.4$ & $40.0\pm6.8$ & $67.9\pm20.3$ & $3.2\pm5.4$ & $2.1\pm1.5$\\
		\cmidrule(lr){2-8}
		& \textbf{Combined} & $\mathbf{2.2\pm0.8}$ & $\mathbf{3.2\pm2.4}$ & $\mathbf{31.9\pm3.6}$ & $\mathbf{49.3\pm16.0}$ & $\mathbf{1.1\pm1.7}$ & $\mathbf{0.7\pm0.6}$\\
		\midrule
		\textbf{Full} & \textbf{All Methods} & $\mathbf{0.2\pm0.4}$ & $\mathbf{1.1\pm2.6}$ & $\mathbf{22.4\pm2.7}$ & $\mathbf{1.6\pm0.9}$ & $\mathbf{0.1\pm0.3}$ & $\mathbf{0.5\pm0.5}$\\
		\bottomrule
	\end{tabular*}

	\caption{\textbf{Counts of unsuccessful adaptations} (mean $\pm$ standard deviation) for LOO on each dataset when using at least one of the given combination of transformations across eight step sizes $\bar{h}=2^{-r}$, $r\in\{1, \ldots, 8\}$, as seen in sixteen simulations of parameter sample size $s=1000$ for ovarian and roaches, eight simulations of $s=500$ for GBSG2 NPES, and sixteen simulations of $s=200$ for GBSG2 QPES. Tested models: logistic regression (LR) and neural network (NN) for ovarian; Poisson regression (PR) and zero-inflated negative binomial regression (NBR) for roaches; neural piecewise exponential survival (NPES) and quilted piecewise exponential survival (QPES) for GBSG2. ``Ours Combined'' uses PMM1/PMM2/PMM3/LL/Var/KL; ``Prior Combined'' uses MM1/MM2/MM3/GRAMIS/MixIS; ``All Methods'' combines both. {\textbf{Lower is better.}}}\label{tab:success}
\end{table}

We scanned eight step sizes $\bar{h}=2^{-r}$ ($r\in\{1, \ldots, 8\}$), evaluating each transformation and the comparison methods (MM1/MM2/MM3/GRAMIS/MixIS) for each $\bar{h}.$ We performed this procedure sixteen times for ovarian and roaches (using samples of size $s=1000$), eight times for GBSG2 NPES ($s=500$), and sixteen times for GBSG2 QPES ($s=200$).
Recall that adaptation is successful if \emph{any} of the considered transformations can reduce $\hat{k} $ to below $0.7$.

\begin{table}[h!]
	\centering
	\small
	\setlength{\tabcolsep}{3pt}
	\begin{tabular*}{\textwidth}{@{\extracolsep{\fill}} l cc cc cc}
		\toprule
		\multirow{2}{*}{\textbf{Method ($\mathcal{O}$ cost)}} & \multicolumn{2}{c}{\textbf{Ovarian} (n=54)} & \multicolumn{2}{c}{\textbf{Roaches} (n=262)} & \multicolumn{2}{c}{\textbf{GBSG2} (n=686)}\\
		\cmidrule(lr){2-3} \cmidrule(lr){4-5} \cmidrule(lr){6-7}
		& LR & NN & PR & NBR & NPES & QPES\\
		\midrule
		PMM1/2, LL ($sp|\bar{h}|$) & $0.96\pm0.12$ & $6.3\pm0.5$ & $3.7\pm0.7$ & $3.2\pm0.3$ & $5.2\pm1.0$ & $2.2\pm0.4$\\
		Var, KL ($snp|\bar{h}|$) & $0.95\pm0.11$ & $6.4\pm0.8$ & $1.2\pm0.2$ & $1.28\pm0.09$ & $2.2\pm0.2$ & $1.9\pm0.2$\\
		PMM3 ($nsp^2|\bar{h}|$) & $3.4\pm0.3$ & $23\pm3$ & $1.2\pm0.2$ & $1.06\pm0.08$ & $5.2\pm1.0$ & $2.2\pm0.4$\\
		\midrule
		MM1/2/3\footnotemark ($sp$; $nsp^2$ for MM3) & $0.19\pm0.02$ & $3.0\pm0.5$ & $0.41\pm0.07$ & $0.50\pm0.04$ & $0.43\pm0.04$ & $0.32\pm0.06$\\
		GRAMIS ($sp \cdot n_{\textrm{it}} n_{\textrm{pr}}$) & $7.9\pm0.9$ & $72\pm13$ & $3.0\pm0.5$ & $11.8\pm0.4$ & $9.8\pm1.9$ & $16\pm2$\\
		MixIS ($snp$) & $0.73\pm0.13$ & $1.4\pm0.1$ & $1.6\pm0.5$ & $0.29\pm0.20$ & $1.5\pm0.2$ & $1.7\pm0.4$\\
		\bottomrule
	\end{tabular*}
	\caption{\textbf{Seconds per adapted observation} (mean $\pm$ std, s/obs; $s=1000$, eight step sizes: $\bar{h}=2^{-r}$ with $r\in\{1,\ldots,8\}$). Here $s$ is the number of posterior samples, $n$ is the number of observations, $p$ is the parameter dimension, $|\bar{h}|$ is the number of step sizes searched, $n_{\textrm{iter}}$ is the number of GRAMIS iterations, and $n_{\textrm{prop}}$ is the number of GRAMIS proposals. Complexity reflects the cost of computing the transformation $Q_i$ and applying it; the KL/Var transformations additionally require evaluating $\pi(\btheta|\calD)$ at $np$ cost per sample, accounting for the $n$ factor. In the MCMC setting, all methods additionally incur an $snp$ shared cost for computing the posterior ratio in the importance weights (Eq.~\ref{eq:importance}), which dominates when $n$ is large. Timings measured after JIT warmup on Jax/CPU (AMD Ryzen AI 395+ Strix Halo). ``--'' indicates timings not yet available. PMM3/MM3 require forming per-observation weighted covariance matrices at $\mathcal{O}(nsp^2 + np^3)$ cost.}\label{tab:timings}
\footnotetext{MM1/2/3 are computed in a single pass; the reported time includes MM3's covariance computation amortized over the batch.}
\end{table}

Table~\ref{tab:success} presents statistics (mean $\pm$ standard deviation) for the number of observations where adaptation fails when using the given combination of methods.
Using all methods in unison generally eliminates the need to refit either the logistic regression or the neural network models for the task of obtaining LOO statistics.
In particular the PMM1/PMM2 methods were highly effective for the neural model.

Fig.~\ref{fig:khat} depicts the minimum $\hat{k}$ per transformation for a representative logistic regression simulation.
PMM1/PMM2 had the most success overall, but there are instances (e.g., observations 23 and 46) where they fail and the gradient flow transformations succeed, illustrating the complementarity of the methods.

\begin{figure}[h!]
	\centering
	\begin{subfigure}{0.8\linewidth}
		\centering
		\includegraphics[width=\linewidth]{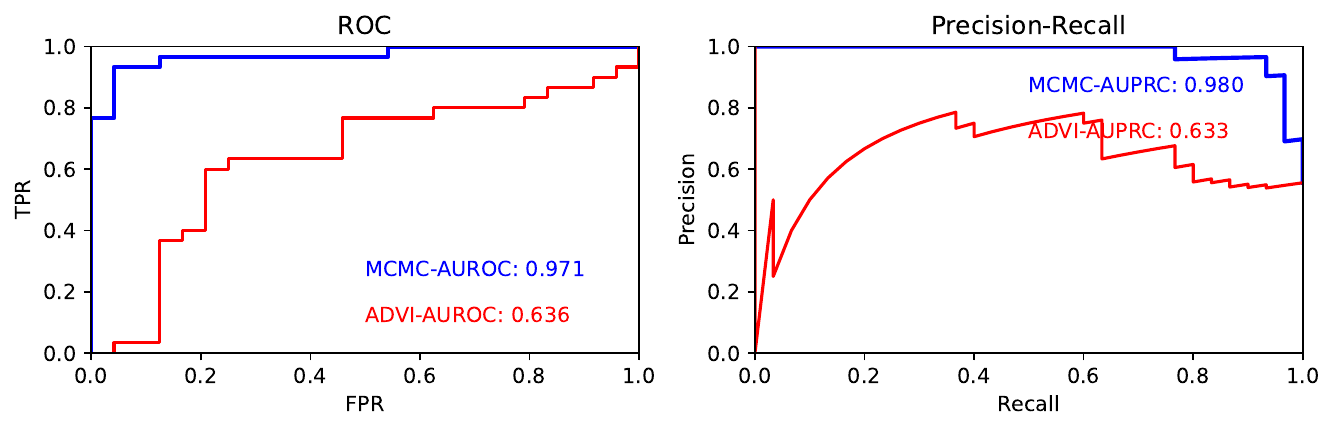}
		\caption{Logistic regression}\label{fig:lr_curves}
	\end{subfigure}\\[0.5em]
	\begin{subfigure}{0.8\linewidth}
		\centering
		\includegraphics[width=\linewidth]{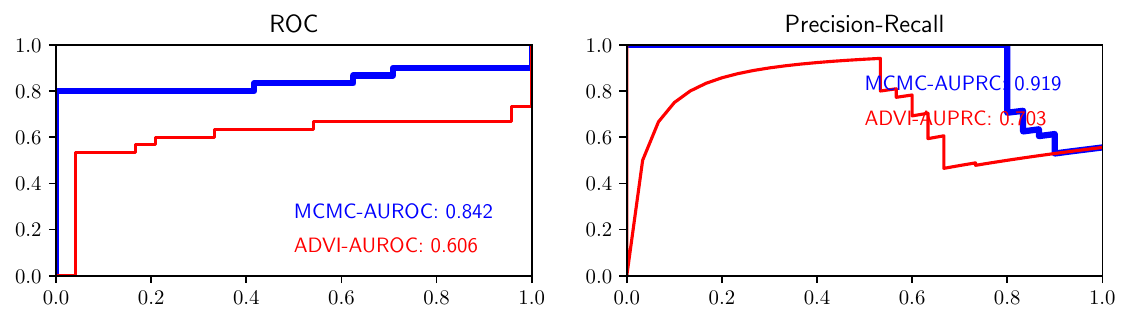}
		\caption{Neural network}\label{fig:relu_curves}
	\end{subfigure}
	\caption{\textbf{LOO ROC curves for ovarian cancer classification models} contrasting MCMC and mean-field ADVI inference. MCMC-inferred models show better generalization in this $p\gg n$ setting.}
	\label{fig:curves}
\end{figure}

The corresponding LOO ROC curves for the ovarian dataset (Fig.~\ref{fig:curves}) show that MCMC-inferred models have better generalization than their mean-field ADVI approximations for both the logistic regression and neural network models, consistent with the expected multicolinearity in this $p\gg n$ problem.

\paragraph{Validation against exact LOO}
To validate the IS-LOO estimates, we computed exact LOO by refitting the ovarian logistic regression model 54 times (once per observation).
Per-observation predictive probabilities from successfully adapted observations ($\hat{k}<0.7$) closely track the ground truth (RMSE $\leq 0.05$; Supplement~\ref{sec:loo_validation}, Fig.~\ref{fig:loo_validation}), and the combined adaptation achieves a LOO-AUC of $0.971$ versus $0.968$ from exact refitting.

\section{Discussion}

We introduced six perturbative transformations -- PMM1/PMM2/PMM3, LL gradient descent, and Var/KL gradient flow steps -- for adapting full-data posteriors to approximate LOO cross validation without refitting.
We derived exact Jacobian determinants for logistic regression and shallow ReLU networks, and described efficient approximations for deeper architectures.
The resulting LOO predictions enable computation of downstream metrics such as ROC/PRC curves at a fraction of the cost of full refitting.

\paragraph{Contrasting and synergizing methods}

Examining Table~\ref{tab:success}, the original MM1/MM2 transformations without split sampling provided limited improvement for the ovarian and roach models, while MM3 was effective for some models (e.g., LR and NPES) but not others.
The generalized PMM transformations with tunable step size were often the strongest individual methods, but no single transformation dominated -- for instance, LL outperformed all PMM variants on the logistic regression.
The strength of the approach lies in this diversity: because the transformations capture different aspects of the LOO perturbation, their combination nearly eliminates adaptation failures even when any individual method leaves many observations unadapted.
The general strategy is to loop through observations and try successive transformations until adaptation is successful, avoiding the cost of refitting.
The contrast between the Poisson and NBR results on the roach data is instructive: the Poisson model's high residual failure count ($22.4$ observations even with all methods combined) reflects model misspecification -- the data exhibits 36\% zeros and substantial overdispersion that Poisson cannot capture.
The NBR model, which accommodates zero-inflation, has the combined methods reducing failures to $1.6\pm0.9$ -- adaptation cannot fully compensate for a poorly specified model.

\paragraph{Limitations and extensions}
The main tradeoff versus model-agnostic PSIS-LOO is model-dependence: one needs gradients of the likelihood and prior.
The Var/KL transformations additionally require the posterior density, which is costly without a variational approximation.
The first-order Jacobian approximation introduces bias scaling as $\mathcal{O}(k^2/n^2)$ at the natural step size $\bar{h}\sim 1/\sqrt{n}$, where $k$ is the effective sparsity of $Q_i$ (Supplement~\ref{sec:jacobianerror}); so, for models with shrinkage priors ($k\ll p$), the error is far smaller than the naive $\mathcal{O}(p/n^2)$ bound suggests.

A natural extension for the gradient flow transformations is to take multiple steps, though at increasing Jacobian cost.
Another is adaptive step size selection (e.g., bisection on $\hat{k}(\bar{h})$) rather than the current grid search, which would reduce the number of candidate evaluations per observation.

\newpage

\bibliographystyle{unsrtnat}
\bibliography{irtvae}

\onecolumn
\title{Perturbative adaptive importance sampling for Bayesian LOO-CV\\(Supplement)}
\maketitle
\setcounter{section}{0}
\setcounter{equation}{0}
\setcounter{figure}{0}
\renewcommand{\thefigure}{S.\arabic{figure}}
\renewcommand{\theequation}{S.\arabic{equation}}
\renewcommand{\thesection}{S.\arabic{section}}

\section{Full-rank partial moment matching (PMM3)}
\label{sec:pmm3}

The PMM1 and PMM2 transformations match the first moment and diagonal second moments respectively of the weighted (leave-one-out) and unweighted distributions, independently for each scalar component.
PMM3 generalizes this to match the full covariance structure.

Let $\bSigma = \frac{1}{s}\sum_{k=1}^s (\btheta_k - \bar\btheta)(\btheta_k - \bar\btheta)^\top$ and $\bSigma_{w,i} = \frac{\sum_k \nu_k (\btheta_k - \bar\btheta_{w,i})(\btheta_k - \bar\btheta_{w,i})^\top}{\sum_k \nu_k}$ denote the unweighted and weighted sample covariance matrices, with Cholesky factorizations $\bSigma = \bL\bL^\top$ and $\bSigma_{w,i} = \bL_{w,i}\bL_{w,i}^\top$.
Define
\begin{equation}
	\bA_i = \bL_{w,i}\bL^{-1} - \bI,
\end{equation}
so that $\bA_i$ captures the rotation and scaling needed to match the weighted covariance.
The PMM3 transformation is
\begin{equation}
	T_{\textrm{PMM}3}(\btheta) = \btheta + \bar{h}\left[\bA_i(\btheta - \bar\btheta) + (\bar\btheta_{w,i} - \bar\btheta)\right].
\end{equation}
At $\bar{h}=1$ this reduces to $\bL_{w,i}\bL^{-1}(\btheta - \bar\btheta) + \bar\btheta_{w,i}$, recovering the MM3 affine map of \citet{paananenImplicitlyAdaptiveImportance2021}.
Since $\bA_i$ is constant in $\btheta$, the Jacobian is exact:
\begin{equation}
	\log|\det J_i| = \log|\det(\bI + \bar{h}\bA_i)| = \sum_{j=1}^p \log|1 + \bar{h}\lambda_j(\bA_i)|,
\end{equation}
where $\lambda_j(\bA_i)$ are the eigenvalues of $\bA_i$.
In particular, the divergence $\nabla\cdot Q_i = \operatorname{tr}(\bA_i)$ is constant, so the first-order Jacobian approximation $\log|J| \approx \bar{h}\operatorname{tr}(\bA_i)$ used in the perturbative framework is exact to all orders.

\paragraph{Computational cost.}
The dominant cost is forming and factorizing the $n$ weighted covariance matrices $\bSigma_{w,i}$, each at $\mathcal{O}(sp^2 + p^3)$ cost, giving $\mathcal{O}(nsp^2 + np^3)$ total.
Computing the full Hessian is feasible when $p$ is moderate but becomes prohibitive for high-dimensional models.
The per-observation cost of the transformation itself is $\mathcal{O}(sp^2)$ for the matrix--vector products.

\section{Derivation of gradient-flow transformations}
\label{sec:variationalprobs}

\subsection{KL Divergence}
\label{sec:KLdivergence}
For convenience, we write the cross-entropy in the forward-transformation form, so that
\begin{align}
	H\left( \pi({\bphi|\calD^{(-i)}}), \pi_{\bphi}(\bphi|\calD)\right) & = -\int \nu_i(\bphi)\pi(\bphi|\calD) \log \frac{\pi(T_i^{-1}(\bphi)|\calD)}{\mathcal{J}_i(T_i^{-1}(\bphi))} \dd\bphi \nonumber                \\
	                                                                   & =-\int \nu_i(T_i(\btheta)) {\pi(T_i(\btheta)|\calD)} \log \frac{\pi(\btheta|\calD)}{\mathcal{J}_i(\btheta)}{\calJ_{T_i}(\btheta)} \dd\btheta.
\end{align}
Now for an arbitrary test function with vanishing boundary conditions, $\xi$,
\begin{align}
	\MoveEqLeft\int \frac{\delta H\left( \pi({\btheta|\calD^{(-i)}}), \pi_{\bphi}(\btheta|\calD)\right)}{\delta T_i}  \xi(\btheta) \dd\btheta  \nonumber                                                                                                                                                                     \\
	 & =- \lim_{\varepsilon\to0}\frac{\dd}{\dd\varepsilon}\int  \nu_i(T_i(\btheta) + \varepsilon\xi(\btheta)) {\pi(T_i(\btheta) + \varepsilon\xi(\btheta)|\calD) }\log \frac{\pi(\btheta|\calD)}{|\nabla T_i(\btheta) + \varepsilon\nabla \xi(\btheta)|}{|\nabla T_i(\btheta) + \varepsilon\nabla\xi|} \dd\btheta  \nonumber \\
	 & = -\int\nabla\left\{ {\pi( \btheta|\calD) } \nu_i( \btheta) \right\}\big|_{\btheta=T_i(\btheta)}{\mathcal{J}_i(\btheta)}\log \frac{\pi(\btheta|\calD)}{\mathcal{J}_i(\btheta)}\cdot\xi(\btheta)\dd\btheta \nonumber                                                                                                   \\
	 & \qquad - \int \nu_i(T_i(\btheta))\pi(T_i(\btheta)|\calD)\left[ {\calJ_{T_i}(\btheta)}\log\frac{ \pi(\btheta|\calD)}{\calJ_{T_i}(\btheta)} -\calJ_{T_i}(
		\btheta)\right]\textrm{tr}\left(\bJ^{-1}_i(\btheta)\nabla\xi(\btheta)\right)\dd\btheta.
\end{align}
Re-arranging the trace term,
\begin{align}
	\textrm{tr} \left[\bJ_i^{-1}\nabla \xi(\btheta) \right] & = \sum_q \sum_r (\bJ_i^{-1})_{qr}(\nabla\xi)_{rq} = \sum_r \sum_q \frac{\partial\xi_r}{\partial\theta_q}(\bJ_i^{-1}(\btheta))_{qr},
\end{align}
and integrating each term by parts,
\begin{align}
	\MoveEqLeft\int \frac{\delta H\left( \pi({\btheta|\calD^{(-i)}}), \pi_{\bphi}(\btheta|\calD)\right)}{\delta T_i}  \xi(\btheta) \dd\btheta  \nonumber                                                                  \\
	 & =-\int\nabla\left\{ {\pi( \btheta|\calD) } \nu_i( \btheta) \right\}\big|_{\btheta=T_i(\btheta)}{\mathcal{J}_i(\btheta)}\log \frac{\pi(\btheta|\calD)}{\mathcal{J}_i(\btheta)}\cdot\xi(\btheta)\dd\btheta \nonumber \\
	 & \qquad + \sum_r \nabla\cdot \int \frac{\partial}{\partial\theta_q}\left\{\nu_i(T_i(\btheta))\pi(T_i(\btheta)|\calD)\left[ {\calJ_{T_i}(\btheta)}\log\frac{ \pi(\btheta|\calD)}{\calJ_{T_i}(\btheta)} -\calJ_{T_i}(
		\btheta)\right](\bJ_i^{-1}(\btheta))_{qr} \right\} \mathbf{e}_r\cdot\xi\dd\btheta.
\end{align}
So, the Euler-Lagrange equation satisfies
\begin{align}
	\MoveEqLeft \frac{\delta \mathcal{H}}{\delta T_i} = -\nabla  ({\pi( \btheta|\calD) } \nu_i( \btheta) ) \big|_{\btheta=T_i(\btheta)}{\mathcal{J}_i(\btheta)}\log \frac{\pi(\btheta|\calD)}{\mathcal{J}_i(\btheta)}  \nonumber \\
	 & \qquad + \sum_r \nabla\cdot \left\{\nu_i(T_i(\btheta))\pi(T_i(\btheta)|\calD)\left[ {\calJ_{T_i}(\btheta)}\log\frac{ \pi(\btheta|\calD)}{\calJ_{T_i}(\btheta)} -\calJ_{T_i}(
	\btheta)\right](\bJ_i^{-1}(\btheta))_{qr} \right\}\mathbf{e}_r \nonumber                                                                                                                                                     \\
	 & =0.
\end{align}

We approximate $T_i$ using the single-step (of size $0<h\ll 1$) forward Euler update under the initial condition $T_i(\btheta, 0)=\btheta,$
\begin{equation}
	T_i(\btheta) \approx T_i(\btheta, 0) - h\frac{\delta H\left( \pi({\btheta|\calD^{(-i)}}), \pi_{\bphi}(\btheta|\calD)\right)}{\delta T_i}\Big|_{T_i(\btheta)=\btheta} = \btheta +h\pi(\btheta|\calD)\nabla\left(\frac{1}{\ell(\btheta|\bx_i, y_i)} \right),
\end{equation}
where we have absorbed the unknown normalizing constant into $h.$

\subsection{Variance}
\label{sec:variance}
The variance of the transformed importance sampling estimator of Eq.~\ref{eq:importance} is specific to the target function.
We examine the variance in computing the expectation of a function $f(\btheta)$ which is related to the variance of the individual element
\begin{align}
	\MoveEqLeft\textrm{Var}\left[ \nu_{i}(\bphi)\calJ_{T_i}(T_i^{-1}(\bphi))\frac{\pi(\bphi|\calD)}{\pi(T_i^{-1}(\bphi)|\calD)} f(\btheta) \right] = \int\left[ \nu_{i}(\bphi) \frac{\pi(\bphi|\calD)}{\pi_{\bphi}(\bphi|\calD)} f(\btheta) \right]^2\pi_{\bphi}(\bphi|\calD)\dd\bphi \nonumber \\
	 & \qquad- \left\{\int \left[ \nu_{i}(\bphi) \frac{\pi(\bphi|\calD)}{\pi_{\bphi}(\bphi|\calD)} f(\btheta) \right] \pi_{\bphi}(\bphi|\calD)\dd\bphi \right\}^2 \nonumber                                                                                                                     \\
	 & = \int \frac{ \left[\nu_{i}(\bphi)\pi(\bphi|\calD) f(\btheta)\right]^2}{\pi_{\bphi}(\bphi|\calD)}\dd\bphi + \textrm{constant relative to }T_i \label{eq:varianceIS}
\end{align}
Note that if one plugs in $f(\btheta) = \mathscr{p}_i(\btheta)$  into Eq.~\ref{eq:varianceIS}, then when $y_i=1$ the overall functional loses dependence on $\ell(\btheta|\bx_i,y_i)$ because $\nu_i$ and $\mathscr{p}_i(\btheta)$ cancel.
For this reason, in order to optimize with respect to the variance of the prediction, one can optimize against the symmetric function
\[ f_i(\btheta) = \mathscr{p}_i(\btheta)^{1-y_i}(1-\mathscr{p}_i(\btheta))^{y_i}.\]

Writing this expression in terms of the forward transformation and doing a change of variables $\dd\bphi = \calJ_{T_i}(\btheta)\dd\btheta,$ we write the functional to minimize
\begin{align}
	\mathcal{V}[T_i] & = \int \frac{g(T_i(\btheta))}{\pi_{\bphi}(T_i(\btheta)|\calD)}\calJ_{T_i}(\btheta)\dd\btheta \nonumber \\
	                 & =  \int \frac{g(T_i(\btheta))}{\pi ( \btheta|\calD)}\calJ^2_i(\btheta)\dd\btheta                       \\
	g(\btheta)       & = (\nu_i(\btheta)\pi(\btheta|\calD)f_i(\btheta))^2.
\end{align}
Computing the first variation of $\mathcal{V}$  with respect to $T_i,$ using Gateaux differentiation,
\begin{align}
	\MoveEqLeft\delta\mathcal{V}
	= \lim_{\varepsilon\to0}\frac{\dd}{\dd\varepsilon}\int \frac{g(T_i(\btheta) + \varepsilon\xi)}{\pi(\btheta|\calD)}\left|\nabla T_i + \varepsilon\nabla\xi\right|^2\dd\btheta \nonumber                                                                               \\
	 & = \int\left\{ \frac{\nabla g(T_i(\btheta))\calJ^2_i(\btheta)}{\pi(\btheta|\calD)}\cdot\xi(\btheta) + 2\frac{g(T_i(\btheta))\calJ^2_i(\btheta)}{\pi(\btheta|\calD)} \textrm{tr}\left(\bJ^{-1}_i(\btheta)\nabla\xi(\btheta)\right) \right\} \dd\btheta \nonumber    \\
	 & = \int\left\{ \frac{\nabla g(T_i(\btheta))\calJ^2_i(\btheta)}{\pi(\btheta|\calD)}  - 2\sum_r \nabla\cdot\left[\frac{g(T_i(\btheta))\calJ^2_i(\btheta)}{\pi(\btheta|\calD)} \bJ^{-1}_i(\btheta)_{,r}\right] \mathbf{e}_r    \right\} \cdot\xi(\btheta) \dd\btheta.
\end{align}
So the Euler-Lagrange equation for minimizing the variance is
\begin{equation}
	\frac{\delta\mathcal{V}}{\delta T_i} =\frac{\nabla g(T_i(\btheta))\calJ^2_i(\btheta)}{\pi(\btheta|\calD)}  - 2\sum_r \nabla\cdot\left[\frac{g(T_i(\btheta))\calJ^2_i(\btheta)}{\pi(\btheta|\calD)} \bJ^{-1}_i(\btheta)_{,r}\right] \mathbf{e}_r  = 0.
	\label{eq:ELvariance}
\end{equation}

For $T_i(\btheta)=\btheta$, the functional derivative is
\begin{align}
	\MoveEqLeft \frac{\delta\mathcal{V}}{\delta T_i}\Big|_{T_i=\btheta} = \frac{\nabla g(\btheta)}{\pi(\btheta|\calD)} - 2\nabla\left( \frac{ g(\btheta)}{\pi(\btheta|\calD)}\right) \nonumber    \\
	 & =\frac{\nabla g(\btheta)}{\pi(\btheta|\calD)} - 2\frac{\nabla g(\btheta)}{\pi(\btheta|\calD)} +2\frac{g\nabla \pi(\btheta)}{\pi^2(\btheta|\calD)} \nonumber                                \\
	 & =-\frac{2\nu_i(\btheta)\pi(\btheta|\calD) f_i(\btheta)\nabla(\nu_i(\btheta)\pi(\btheta) f_i(\btheta))}{\pi(\btheta|\calD) } + 2\frac{g\nabla \pi(\btheta)}{\pi^2(\btheta|\calD)} \nonumber \\
	 & =-2 \nu_i(\btheta)  f_i(\btheta)\nabla(\nu_i(\btheta)\pi(\btheta|\calD) f_i(\btheta)) + 2\nu_i^2(\btheta)f_i^2(\btheta)\nabla\pi(\btheta|\calD) \nonumber                                  \\
	 & = -2\nu_i(\btheta)f_i(\btheta)\pi(\btheta|\calD)\nabla\big[\nu_i(\btheta)f_i(\btheta)\big]
\end{align}

\section{Sigmoidal models}

\subsection{Jacobians of sigmoidal model transformations}
\label{sec:sigmoidjacobian}

For sigmoidal classification models the Jacobians of the KL and variance descent transformations take the form
\begin{align}
	\MoveEqLeft \bJ_{T_i^{\textrm{KL/Var}}}(\btheta) = \bI + \Bigg\{h(-1)^{y_i}{\pi(\btheta|\calD)}  e^{(1 + 1_{\textrm{Var}})\mu_i(\btheta)(1-2y_i)} \nonumber \\
	 & \qquad\times\Big\{  \nabla \nabla\mu_i
	+ \left[\nabla\log\pi(\btheta|\calD) + (1+1_{\textrm{Var}}) (1-2y_i)\nabla\mu_i \right] (\nabla\mu_i)^\intercal \Big\}  \Bigg\}
\end{align}
where
$
	\nabla\log\pi(\btheta|\calD) = \nabla\log\pi(\btheta)  + \sum_j\left( y_j(1-\sigma(\mu_j))-(1-y_j)\sigma(\mu_j) \right) \nabla \mu_j(\btheta),
$
and
$\pi(\btheta)$ is the prior.

\subsection{Logistic regression Jacobian determinants}
\label{sec:lrjacobian}

For logistic regression, $\mu_i(\btheta)=\bx_i^\intercal\bbeta$, so $\nabla\mu_i = \bx_i$ and $\nabla\nabla\mu = \mathbf{0}.$ The Jacobian of $Q_i$ is rank-one with a single non-zero eigenvalue, yielding the exact determinant:
\begin{align}
	\calJ_{T^{\textrm{KL/Var}}_i}(\btheta) = \Big| 1 + h(-1)^{y_i}{\pi(\btheta|\calD)}  e^{(1+1_{\textrm{Var}})\mu_i(\btheta)(1-2y_i)} \bx_i^\intercal\left[\nabla\log\pi(\btheta|\calD) + (1+1_{\textrm{Var}})(1-2y_i)\bx_i \right]  \Big|.
\end{align}

\subsection{Gradient and Hessian of the likelihood}

For these models one may use the chain rule to write the gradient,
\begin{align}
	\nabla \ell(\btheta | \bx, y)      & = \ell(\btheta | \bx, y)\nabla\log\ell(\btheta | \bx, y)                    \\
	\nabla \log \ell(\btheta | \bx, y) & = \underbrace{[ y(1-\sigma(\mu))-(1-y) \sigma(\mu)]}_{(\log\ell)'}\nabla\mu
	\label{eq:sigmoidgradient}
\end{align}
and Hessian
\begin{align}
	\MoveEqLeft\nabla\nabla\ell(\btheta | \bx, y) = \ell(\btheta | \bx, y)\nabla\nabla\log\ell(\btheta | \bx, y) \nonumber \\
	 & + \ell(\btheta | \bx, y)\nabla\log\ell(\btheta | \bx, y) \nabla\log\ell(\btheta | \bx, y)   \label{eq:Hell}         \\
	\MoveEqLeft\nabla\nabla\log \ell(\btheta | \bx, y)=  [ y(1-\sigma(\mu))-(1-y) \sigma(\mu)]\nabla\nabla\mu  \nonumber   \\
	 & \underbrace{-\sigma(\mu)(1-\sigma(\mu))}_{(\log\ell)''}\nabla\mu \nabla\mu
	\label{eq:Hlogell}
\end{align}
for the likelihood function as a function of the gradient and Hessian of $\mu$.

\subsection{Shallow Bayesian (ReLU) neural networks}
\label{sec:reluexample}

Bayesian ReLU-nets~\citep{leeConsistencyPosteriorDistributions2000,ghoshModelSelectionBayesian2017,choiRelativeExpressivenessBayesian2018,kristiadiBeingBayesianEven2020,bhadraHorseshoeRegularizationMachine2019} are piecewise linear~\citep{sudjiantoLinearIterativeFeature2021,wangEstimationComparisonLinear2022,montufarNumberLinearRegions2014,sudjiantoUnwrappingBlackBox2020} regression models.
Being locally linear, these models have block-sparse Hessians and are also amenable to some limited degree of interpretability~\citep{sudjiantoUnwrappingBlackBox2020,changInterpretableNotJust2024}.
One may write an $L$-layer ReLU Bayesian neural network recursively, $y_i | \mu_i\sim \textrm{Bernoulli}(\sigma(\mu_i) )$, where
\begin{align}
	\mu_i | \bW_L, b_L, \bz^{(i)}_{L-1} & = \mu(\bx_i) =  \bW_L a(\bz^{(i)}_{L-1}) + b_L \nonumber \\
	\bz_k|\bz^{(i)}_{k-1}, \bb_k, \bW_k & = \bW_k a(\bz^{(i)}_{k-1}) + \bb_k \nonumber             \\
	\bz^{(i)}_1 |\bW_1, \bx_i           & = \bW_1 \bx_i,\label{eq:relunet}
\end{align}
where
$a$ is the ReLU activation function.
Within the parameterization of Eq.~\ref{eq:relunet} we absorbed the initial first-layer bias into the transformation $\bW_1,$ by assuming that $\bx$ has a unit constant component, as is the convention in  regression.

The Hessian matrix of $\mu$, while non-zero, is sparse because all of the following identities hold:
$
	\nabla_{\bb_k}\nabla_{\bb_j} \mu = \mathbf{0} \ \ \forall j, k, \quad
	\nabla_{\bW_k} \nabla_{\bW_k}\mu =\mathbf{0}  \ \ \forall k,  \quad
	\nabla_{\bW_k}\nabla_{\bb_j}  \mu =\mathbf{0}  \ \forall j\geq k.
$
For this reason, the Jacobian determinant approximation of Eq.~\ref{eq:Japprox} can ignore the model Hessian entirely.
However, in the case of one hidden layer we exploit the Hessian's structure to provide explicit exact expressions for $\calJ_{(\cdot)}.$

Sigmoidal classification models are governed by the equations
$\mu = \bW_2 a(\bz_1) + b_2$ and
$\bz_1 = \bW_1\bx$,
where $\bW_2 \in \mathbb{R}^{1\times d},$ $b_2\in\mathbb{R}$, $\bW_1 \in \mathbb{R}^{d\times p},$ $\bb_1\in\mathbb{R}^d.$
The model has the first-order derivatives
$
	{\partial_{({W}_{2})_{1i}}}\mu = a(z_{1})_{i} $,
$ {\partial_{({W}_{1})_{ij}}}\mu = (W_{2})_{1i}a'((z_{1})_{i}) x_j$,
${\partial_{b_2}}\mu = 1. $
The only non-zero components of the Hessian matrix for $\mu$ are the mixed partial derivatives
\begin{equation}
	\frac{\partial^2\mu}{\partial{(W_{1})_{jk}}\partial ({W}_{2})_{ 1j}} = a'((z_{1})_j)x_k. 
	\label{eq:partialH}
\end{equation}
For models with a single hidden layer, the Hessian matrix of $\mu$ has a particular block structure that can be exploited (see~\ref{sec:reluonedecomposition} for derivations) in order to find explicit expressions for its $2d$ non-zero eigenvalues, for $k\in\{1,2\ldots,d\}$,
\begin{align}
	\lambda_k^{\pm} = \pm \left[{\sum_j a'((z_1)_k)x_j^2} \right]^{1/2},
	\label{eq:reluoneeigen}
\end{align}
and associated eigenvectors
\begin{equation}
	\bv_k^\pm = \left(
	\begin{matrix}
			\tilde{\bu}_k/\sqrt{2|\bu_k|^2} & \pm\mathbf{e}_k/\sqrt{2} & 0
		\end{matrix}\right)^\intercal, \quad
	\textrm{where}\quad
	\tilde\bu_k = \big(
	\overbrace{0 \quad \ldots \quad  0}^{(k-1)p \textrm{ zeros}} \quad  \bu_k^\intercal \quad  \overbrace{0 \quad  \ldots \quad  0}^{(d-k)p \textrm{ zeros}}
	\big)^\intercal,
\end{equation}
and
$
	\bu_k = a'( (z_1)_k)\bx.$
To compute the overall transformation Jacobians, one can then apply rank-one updates to $\nabla\nabla\mu$ -- a process that is aided by projecting the model gradients into the eigenspace of the model Hessian (see~\ref{sec:reluone} for derivations).

\subsection{ReLU with one hidden layer}
\label{sec:reluone}

Ordering the parameters $\textrm{vec}(\bW_1), \bW_2, b_2$ (i.e., $(W_1)_{11}, \ldots, (W_1)_{dp}, (W_2)_1,\ldots, (W_2)_d, b_2$), the Hessian of $\mu$ takes the form
\begin{equation}
	\nabla\nabla^\intercal\mu =  \begin{pNiceArray}{ccc}
		\bigzero & \bM & \bigzero \\
		\bM^\intercal & \bigzero  & \bigzero \\
		\bigzero  & \bigzero  & 0
	\end{pNiceArray},\label{eq:HM}
\end{equation}
where $\bM$ encodes the mixed partial derivatives of $\mu$ with respect to the elements of $\bW_1$ and $\bW_2,$
\begin{align}
	\frac{\partial^2\mu}{\partial (W_1)_{ij} (W_2)_{1k}} = \delta_{ik} a'((z_1)_i) x_j.
\end{align}
The $dp\times d$ matrix $\bM$ takes the form
\begin{equation}
	\bM = \left(
	\begin{matrix}
		\bu_1    & \bigzero & \bigzero & \cdots & \bigzero \\
		\bigzero & \bu_2    & \bigzero & \cdots & \bigzero \\
		\bigzero & \bigzero & \bu_3    & \ddots & \vdots   \\
		\vdots   & \vdots   & \ddots   & \ddots & \vdots   \\
		\bigzero & \cdots   & \cdots   & \cdots & \bu_d
		\label{eq:M}
	\end{matrix}
	\right)
\end{equation}
where each column vector
\begin{equation}
	\bu_k = \left(
	\begin{matrix}
		a'( (z_1)_k) x_1 & a'( (z_1)_k) x_2 & \cdots & a'( (z_1)_k) x_p
	\end{matrix}
	\right)^\intercal = a'( (z_1)_k)\bx
\end{equation}
is potentially sparse.

\subsubsection{Hessian decomposition}
\label{sec:reluonedecomposition}
Suppose that $\lambda$ is an eigenvalue of $\nabla\nabla^\intercal\mu$, corresponding to the eigenvector $\boldsymbol\psi = \left(\begin{matrix}  \boldsymbol\psi_1^\intercal & \boldsymbol\psi_2^\intercal & \boldsymbol\psi_3^\intercal \end{matrix}\right)^\intercal$.
Then,
\begin{align}
	\bM \boldsymbol\psi_2          & = \lambda \boldsymbol\psi_1 \label{eq:psi2_1} \\
	\bM^\intercal\boldsymbol\psi_1 & = \lambda \boldsymbol\psi_2 \label{eq:psi1_2} \\
	\boldsymbol\psi_3              & = \mathbf{0}.
\end{align}
Left-multiplying Eq.~\ref{eq:psi2_1} by $\bM^\intercal$ and Eq.~\ref{eq:psi1_2} by $\bM$, one finds that
\begin{align}
	\bM^\intercal \bM \boldsymbol\psi_2 & = \lambda \bM^\intercal\psi_1 = \lambda^2 \boldsymbol\psi_2\label{eq:psi2}                \\
	\bM\bM^\intercal\boldsymbol\psi_1   & = \lambda \bM\bM^\intercal \boldsymbol\psi_2 = \lambda^2\boldsymbol\psi_1.\label{eq:psi1}
\end{align}
Eq.~\ref{eq:psi2} implies that solutions for  $\boldsymbol\psi_2$ are eigenvectors of $\bM^\intercal\bM$ and the solutions for $\boldsymbol\psi_1$ are eigenvectors of $\bM\bM^\intercal.$

The matrix $\bM^\intercal\bM$ is the diagonal matrix $\textrm{diag}(|\bu_k|^2)$, so the non-zero eigenvalues are each of $\pm |u_k|,$ and the solutions for $\boldsymbol\psi_2$ are proportional to elements of the $d$ dimensional standard basis.

It is easy to see that the eigenvectors of $\bM\bM^T$, which we denote  $\tilde\bu_k$, are  related to the column vectors of $\bM$,
\begin{equation}
	\tilde\bu_k = \big(
	\overbrace{0 \quad \ldots \quad  0}^{(k-1)p \textrm{ zeros}} \quad  \bu_k^\intercal \quad  \overbrace{0 \quad  \ldots \quad  0}^{(d-k)p \textrm{ zeros}}
	\big)^\intercal.
\end{equation}
Overall, each eigenvector for $\nabla\nabla^\intercal\mu$ takes the form $\big(\tilde{\bu}_k^\intercal\quad \gamma \mathbf{e}_k^\intercal \quad 0\big)^\intercal$ for some constant $\gamma$.
By Eq.~\ref{eq:psi2},
\begin{align}
	\bM \gamma \mathbf{e}_k = \gamma \tilde{\bu}_k= \lambda \tilde{\bu}_k ,\nonumber
\end{align}
so $\gamma = \lambda.$ Hence, by spectral theorem, we may write the decomposition for $\nabla\nabla^\intercal\mu,$
\begin{equation}
	\nabla\nabla^\intercal\mu = \sum_{k=1}^d \left[
		|\bu_k|\,\bv_{k+}\bv_{k+}^\intercal - |\bu_k|\,\bv_{k-}\bv_{k-}^\intercal
		\right],
\end{equation}
where
\begin{equation}
	\bv_{k\pm} = \left(
		\begin{matrix}
			\tilde{\bu}_k/\sqrt{2|\bu_k|^2} \\ \pm\mathbf{e}_k/\sqrt{2} \\ 0
		\end{matrix}\right).
\end{equation}

\subsubsection{Overall model Jacobians}
To assemble the Jacobian of the overall transformation, we need to diagonalize the gradient outer product simultaneously with $\bM.$ The gradient terms follow
\begin{align}
	\frac{\partial\mu}{\partial b_2}        & = 1                           \\
	\frac{\partial\mu}{\partial (W_2)_{1k}} & = a((z_1)_k)                  \\
	\frac{\partial\mu}{\partial (W_1)_{jk}} & = (W_{2})_{j} a'((z_1)_j) x_k
\end{align}
so that
\begin{equation}
	\nabla\mu = \left(
	\begin{matrix}
		(W_2)_1\bu_1^\intercal & (W_2)_2\bu_2^\intercal & \cdots & (W_2)_d\bu_d^\intercal & a((z_1)_1) & a((z_1)_2) & \cdots & a((z_1)_d) & 1
	\end{matrix}
	\right)^\intercal
\end{equation}

We then decompose the gradient into the eigenspace of the Hessian,
\begin{align}
	\nabla\mu & = \sum_{k=1}^d \left[\frac{  (W_2)_k |\bu_k| + a((z_1)_k)  }{\sqrt{2}}\bv_{k+} + \frac{ (W_2)_k |\bu_k| - a((z_1)_k) }{\sqrt{2}}\bv_{k-}\right] +  \underbrace{\left(
	\begin{matrix}
			0 & \cdots & 0 & 1
		\end{matrix}\right)^\intercal}_{\mathbf{e}_{dp+d + 1}} \nonumber                                                                                                                  \\
	          & = \mathbf{V}\mathbf{c}
	\label{eq:gradientdecomp}
\end{align}
where $\mathbf{V}$ is the matrix where the columns are the eigenvectors
\begin{align}
	\mathbf{V} & = \left(
	\begin{matrix}
			\mathbf{v}_{1+} & \mathbf{v}_{1-} & \bv_{2+} & \bv_{2-} & \cdots & \bv_{d+} & \bv_{d-} & \mathbf{e}_{dp+d+1}
		\end{matrix}
	\right),
\end{align}
and $\mathbf{c}$ is a column vector corresponding to the coefficients in Eq.~\ref{eq:gradientdecomp}.

The overall model Jacobians for the KL/Var transformations take the form
\begin{align}
	J(\btheta) & = \bI + \alpha(\btheta)\nabla\nabla^\intercal\mu + \beta(\btheta)\left( \mathbf{r} + \nabla\mu \right) \nabla^\intercal\mu,
\end{align}
where $\alpha, \beta$  are scalars dependent on $\btheta$, and $\mathbf{D}_V$ is a diagonal matrix of the eigenvalues of the Hessian of $\mu,$ and $\mathbf{r}$ is linearly independent of all columns of $\bV$.
The determinant of this matrix is the rank-one update,
\begin{align}
	|J(\btheta)| & = \left| \bI + \alpha(\btheta)\bV\mathbf{D}_V\bV^\intercal \right|\left( 1 + \beta \nabla\mu^\intercal(\bI + \alpha(\btheta)\bV\mathbf{D}_V\bV^\intercal)^{-1}\left(\mathbf{r}+\nabla\mu \right)\right) \nonumber                                        \\
	             & =\left[\prod_{k=1}^d \left( 1+ \alpha(\btheta)\lambda_k^+\right)\left( 1+ \alpha(\btheta)\lambda_k^-\right)\right] \left( 1 + \beta \nabla\mu^\intercal(\bI + \alpha(\btheta)\bV\mathbf{D}_V\bV^\intercal)^{-1}\left(\mathbf{r}+\nabla\mu \right)\right)
\end{align}

Putting it all together, the Jacobians of the transformation are, for the log-likelihood descent transformation,
\begin{align}
	\mathbf{I} - h\nabla\nabla^\intercal\log\ell & = \mathbf{I} - h\Bigg\{ [ y(1-\sigma(\mu))-(1-y) \sigma(\mu)]\overbrace{\left[\sum_{k=1}^d\left( |\bu_k|\bv_{k+}\bv_{k+}^\intercal- |\bu_k| \bv_{k-}\bv_{k-}^\intercal\right)  \right]}^{\nabla\nabla^\intercal\mu} \nonumber \\
	                                             & \qquad -\sigma(\mu)(1-\sigma(\mu)) \bV\mathbf{c}\mathbf{c}^\intercal\bV^\intercal\Bigg\}.\label{eq:THess}
\end{align}
To compute the determinant of Eq.~\ref{eq:THess} we first make two observations.
First, two similar matrices have the same determinant, so that
\begin{align}
	\MoveEqLeft\textrm{det}\left[ \mathbf{I} - h\nabla\nabla^\intercal\log\ell \right] \nonumber\\
	& = \textrm{det}\Big[\mathbf{I} - h\big[ (\log\ell)' \bV \textrm{diag}(\lambda_i) \bV^\intercal -(\log\ell)'' \bV\mathbf{c}\mathbf{c}^\intercal\bV^\intercal\big] \Big] \nonumber \\
	& =\textrm{det}\Big[\underbrace{\mathbf{I} - h (\log\ell)' \textrm{diag}(\lambda_i)}_{\textrm{diagonal}} +h \underbrace{ (\log\ell)''  \mathbf{c}\mathbf{c}^\intercal }_{\textrm{rank-one}}\Big],\label{eq:intermediatedet}
\end{align}
where $(\log\ell)' = y(1-\sigma(\mu))-(1-y)\sigma(\mu)$ and $(\log\ell)'' = -\sigma(\mu)(1-\sigma(\mu))$ as defined in Eq.~\ref{eq:sigmoidgradient}.
and second that the non-diagonal elements of the matrix in Eq.~\ref{eq:intermediatedet} are rank-one so that
\begin{align}
	\calJ_{T^\textrm{LL}}(\btheta) & = \prod_{j=1}^d\left(1-h^2 [ y(1-\sigma(\mu))-(1-y) \sigma(\mu)]^2|\bu_j|^2) \right)\nonumber                                                                               \\
	                               & \times\Bigg(1+h \sigma(\mu)(1-\sigma(\mu))\Bigg[\sum_{k=1}^d\Bigg[\frac{[ (W_2)_k |\bu_k| + a((z_1)_k) ]^2 }{2[1-h [ y(1-\sigma(\mu))-(1-y) \sigma(\mu)]|\bu_k|]} \nonumber \\
	                               & \qquad\qquad+ \frac{ [(W_2)_k |\bu_k| - a((z_1)_k)]^2 }{2[1+h [ y(1-\sigma(\mu))-(1-y) \sigma(\mu)]|\bu_k|]}\Bigg] + 1\Bigg]\Bigg).
\end{align}

\section{Step size analysis and Jacobian approximation error for KL/Var}
\label{sec:jacobianerror}

\subsection{Optimal step size from influence function analysis}

Under standard regularity conditions (Bernstein--von Mises), the posterior concentrates around the MLE with covariance $\Sigma \approx J(\btheta_0)^{-1}/n$, where $J$ is the Fisher information matrix.
The shift in the posterior mean from removing observation $i$ follows the empirical influence function:
\begin{equation}
	\Delta\btheta^* = \hat\btheta - \hat\btheta_{(-i)} \approx J^{-1}\nabla\log\ell(\hat\btheta|\bd_i)/n = \Sigma\nabla\log\ell(\hat\btheta|\bd_i).
\end{equation}
In posterior standard deviation units ($\sigma_\alpha = \sqrt{\Sigma_{\alpha\alpha}}$), for a typical (non-outlier) observation with $\mathbb{E}[(\nabla_\alpha\log\ell_i)^2] = J_{\alpha\alpha}$:
\begin{equation}
	|\Delta\theta^*_\alpha|/\sigma_\alpha \sim \frac{|\nabla_\alpha\log\ell_i|}{\sqrt{nJ_{\alpha\alpha}}} \sim \frac{1}{\sqrt{n}}.
\end{equation}
Since the step size normalization (Eq.~\ref{eq:hi} in the main text) ensures the transformation step is at most $\bar{h}$ posterior standard deviations, the natural choice is $\bar{h}\sim 1/\sqrt{n}$ to match the LOO perturbation magnitude.

For high-leverage observations, the constant in $\Delta\theta^*/\sigma$ can be substantially larger than $1/\sqrt{n}$ -- these are precisely the observations where $\hat{k}>0.7$ and adaptation is needed.
When $n$ is small relative to $p$, the BvM approximation may not hold and the $1/\sqrt{n}$ scaling breaks down.

\subsection{Optimal step size from the cross-entropy functional}

The KL transformation arises from a single forward Euler step along the gradient flow $\partial T_i/\partial t = -\delta H/\delta T_i$ starting from the identity $T_i=\textrm{id}$:
$T_i(\btheta) \approx \btheta + h Q_i(\btheta)$ where $Q_i = -\delta H/\delta T_i|_{T_i=\textrm{id}}.$
Rather than treating $h$ as a free parameter, one can determine the optimal step size by performing a line search along the direction $Q_i$.
Expanding the cross-entropy to second order in $h$:
\begin{equation}
	H[\textrm{id} + hQ_i] \approx H[\textrm{id}] - h\|Q_i\|^2_{L^2} + \frac{h^2}{2}\left\langle Q_i, \frac{\delta^2 H}{\delta T_i^2}\cdot Q_i\right\rangle,
\end{equation}
where the linear term uses $Q_i = -\delta H/\delta T_i$.
Minimizing this quadratic gives the optimal step size
\begin{equation}
	h^* = \frac{\|Q_i\|^2_{L^2}}{\left\langle Q_i, \frac{\delta^2 H}{\delta T_i^2}\cdot Q_i\right\rangle}. \label{eq:hstar}
\end{equation}
The denominator involves the second variation of the cross-entropy evaluated at the identity, which captures the curvature of $H$ along the descent direction.
For well-specified models, $\delta^2 H/\delta T_i^2$ scales with the Fisher information and the sample size $n$, so $h^*\sim 1/n$ in parameter space, which after the normalization by posterior standard deviations (Eq.~\ref{eq:hi}) corresponds to $\bar{h}^*\sim 1/\sqrt{n}$, consistent with the influence function analysis.

In practice, computing $\delta^2 H/\delta T_i^2$ is expensive, so the grid search over $\bar{h}$ centered around $1/\sqrt{n}$ serves as a practical approximation to the line search in Eq.~\ref{eq:hstar}.
The same analysis applies to the variance functional $\mathcal{V}[T_i]$.

\subsection{Jacobian approximation error}

After the step size normalization, define the normalized vector field $\hat{Q}_{i,\alpha} = Q_{i,\alpha}/(\sigma_\alpha\|\tilde{Q}_i\|_\infty)$ where $\|\tilde{Q}_i\|_\infty = \max_{s,\alpha}|Q_{i,\alpha}(\btheta_s)/\sigma_\alpha|$, so that $|\hat{Q}_{i,\alpha}|\leq 1$ for all components and samples.
The transformation Jacobian becomes
\begin{equation}
	\bJ_{T_i} = \bI + \bar{h}\,\mathrm{diag}(\boldsymbol\sigma)\,\nabla\hat{Q}_i.
\end{equation}
The eigenvalues of $\bJ_{T_i}$ are $1 + \bar{h}\mu_j$ where $\mu_j$ are eigenvalues of $\mathrm{diag}(\boldsymbol\sigma)\nabla\hat{Q}_i$.

The normalized field $\hat{Q}_i$ is bounded ($|\hat{Q}_{i,\alpha}|\leq 1$) and, for models with shrinkage priors, \emph{sparse}: only $k\ll p$ components are significantly nonzero.
If $\hat{Q}_{i,\alpha}(\btheta)\approx 0$ throughout the posterior support for component $\alpha$, then the $\alpha$-th row of $\nabla\hat{Q}_i$ is also approximately zero.
Consequently, $\nabla\hat{Q}_i$ has effective rank $\leq k$, and only $k$ eigenvalues $\mu_j$ differ from zero.
Each nonzero eigenvalue satisfies $|\mu_j|\sim \sigma_\alpha|\partial_\beta\hat{Q}_{i,\alpha}| = \mathcal{O}(1/\sqrt{n})$ since $\sigma\sim 1/\sqrt{n}$ (BvM) and the derivatives of the bounded function $\hat{Q}$ are $\mathcal{O}(1)$ per posterior standard deviation.

The log-determinant and its first-order approximation are
\begin{align}
	\log\det(\bJ_{T_i}) &= \sum_{j=1}^k \log(1 + \bar{h}\mu_j) = \bar{h}\sum_j\mu_j - \tfrac{\bar{h}^2}{2}\sum_j\mu_j^2 + \mathcal{O}(\bar{h}^3), \\
	\log|1 + \bar{h}\,\mathrm{tr}(\mathrm{diag}(\boldsymbol\sigma)\nabla\hat{Q}_i)| &= \bar{h}\sum_j\mu_j - \tfrac{\bar{h}^2}{2}\left(\sum_j\mu_j\right)^2 + \mathcal{O}(\bar{h}^3).
\end{align}
The error in log-space is therefore
\begin{equation}
	\Delta\log\calJ = \bar{h}^2\sum_{j<j'}\mu_j\mu_{j'} + \mathcal{O}(\bar{h}^3), \label{eq:jacoberror}
\end{equation}
which involves at most $\binom{k}{2}$ cross-eigenvalue terms, each of size $\mathcal{O}(1/n)$.
When $k=1$ (e.g., LL on any model whose log-likelihood Hessian is rank-one), the error vanishes exactly -- the first-order approximation is exact.
For general $k$, the error is $\mathcal{O}(\bar{h}^2 k^2/n)$, controlled by the effective sparsity of the transformation rather than the ambient parameter dimension $p$.

For the ovarian horseshoe logistic regression ($p=1057$, $n=54$), the horseshoe prior shrinks most coefficients toward zero, yielding effective sparsity $k\sim 5$--$10$ active features.
The naive bound $\mathcal{O}(\bar{h}^2 p/n)\approx 0.36$ at $\bar{h}=1/\sqrt{n}$ overstates the error by orders of magnitude; the sparsity-corrected bound $\mathcal{O}(\bar{h}^2 k^2/n)$ is $\mathcal{O}(10^{-3})$.
In any case, exact Jacobians are used for the logistic regression and ReLU models in our experiments.

In practice, the $\hat{k}$ diagnostic provides a post-hoc check: if the Jacobian approximation error corrupts the importance weights, $\hat{k}$ will exceed the threshold and that step size will not be selected.

\subsection{Propagation of Jacobian error through self-normalized importance sampling}
\label{sec:jacobianpropagation}

Suppose the approximate Jacobian introduces a per-sample multiplicative error $\epsilon_k$, so that the computed importance weights are $\hat\eta_{ik} = \eta_{ik}(1+\epsilon_k)$ where $\eta_{ik}$ are the exact weights (Eq.~\ref{eq:importance}).
The self-normalized IS estimator with approximate weights is
\begin{equation}
	\hat{\mathbb{E}}[f] = \frac{\sum_k \eta_{ik}(1+\epsilon_k) f(\bphi_k)}{\sum_k \eta_{ik}(1+\epsilon_k)}.
\end{equation}
Expanding to first order in $\epsilon_k$ around $\epsilon_k=0$:
\begin{equation}
	\hat{\mathbb{E}}[f] \approx \bar{f} + \frac{\sum_k \eta_{ik}(\epsilon_k - \bar\epsilon)(f(\bphi_k) - \bar{f})}{\sum_k \eta_{ik}},
\end{equation}
where $\bar{f} = \sum_k \eta_{ik}f(\bphi_k)/\sum_k\eta_{ik}$ is the exact estimator and $\bar\epsilon = \sum_k\eta_{ik}\epsilon_k/\sum_k\eta_{ik}$ is the importance-weighted mean error.
The bias is therefore
\begin{equation}
	\hat{\mathbb{E}}[f] - \bar{f} \approx \textrm{Cov}_{\eta}\left[\epsilon_k, f(\bphi_k)\right], \label{eq:jacobbias}
\end{equation}
i.e., the covariance between the per-sample Jacobian error and the target function under the importance-weighted distribution.

Two consequences follow:
\begin{enumerate}
\item If $\epsilon_k$ is approximately constant across samples (as for the LL transformation, where $Q_i$ does not depend on $\pi(\btheta|\calD)$), then $\epsilon_k \approx \bar\epsilon$ and the bias in Eq.~\ref{eq:jacobbias} vanishes -- the self-normalization cancels a uniform multiplicative error exactly.
\item For the KL/Var transformations, $\epsilon_k$ depends on $\pi(\btheta_k|\calD)$ and thus varies systematically across samples: modal samples have larger Jacobian errors than tail samples. The bias is then $\mathcal{O}(\bar{h}^2 k^2/n)$ times the correlation between the Jacobian error pattern and the target function $f$, where $k$ is the effective sparsity of $Q_i$.
\end{enumerate}

\section{GRAMIS Implementation Details}
\label{sec:gramis_implementation}

The GRAMIS algorithm~\citep{elviraGradientbasedAdaptiveImportance2022} adapts a population of $K$ Gaussian proposals $\{q_k = \mathcal{N}(\bmu_k, \mathrm{diag}(\bsigma_k^2))\}_{k=1}^K$ via gradient ascent on the target density.
For LOO-CV, drawing fresh samples from the mixture $q_{\mathrm{mix}} = K^{-1}\sum_k q_k$ and reweighting to the LOO posterior $\pi(\btheta|\calD^{(-i)})$ suffers from severe weight degeneracy when $n \gg p$, because the full posterior is sharply concentrated around the mode and the Gaussian proposals cannot match it closely enough.

To address this, we employ a \emph{defensive mixture} approach for models with $p \leq 50$ parameters.
The original $s$ posterior samples $\{\btheta_s\}$ are pooled with $M$ fresh samples $\{\bphi_k\}$ drawn from the adapted mixture $q_{\mathrm{mix}}$.
Importance weights are computed against the defensive mixture density
\begin{equation}
	q_{\mathrm{DM}}(\btheta) = \alpha\,\hat{p}(\btheta|\calD) + (1-\alpha)\,q_{\mathrm{mix}}(\btheta),
\end{equation}
where $\hat{p}(\btheta|\calD) = \mathcal{N}(\bar\btheta, \mathrm{diag}(\hat\bsigma^2))$ is a Gaussian fit to the posterior samples and $\alpha=0.5$.
For each sample (whether from the posterior or the mixture), the importance weight targeting $\pi(\btheta|\calD^{(-i)})$ is
\begin{equation}
	w(\btheta) \propto \frac{\hat{p}(\btheta|\calD)}{\ell_i(\btheta)\,q_{\mathrm{DM}}(\btheta)},
\end{equation}
followed by PSIS smoothing on the combined pool.

For high-dimensional models ($p > 50$), the diagonal Gaussian approximation $\hat{p}$ becomes unreliable (the log-density is dominated by the normalizing constant, making all weights nearly uniform and $\hat{k}$ artificially low).
In this regime, GRAMIS falls back to the standard mode using only fresh mixture samples with weights $w(\bphi) \propto \pi(\bphi|\calD^{(-i)}) / q_{\mathrm{mix}}(\bphi)$.

The proposal covariance is adapted using the exact Hessian diagonal of $\log\ell_i$ (computed via forward-over-reverse automatic differentiation), yielding the perturbative update $\sigma_{-i,j}^2 = \sigma_j^2 / (1 + \sigma_j^2 H_{i,jj})$, where $H_{i,jj} = \partial^2\log\ell_i/\partial\theta_j^2 \leq 0$ for log-concave likelihoods.

\section{Supplemental Results}
\label{sec:suppresults}

\subsection{Validation against exact LOO}
\label{sec:loo_validation}

\begin{figure}[h!]
	\centering
	\includegraphics[width=0.9\textwidth]{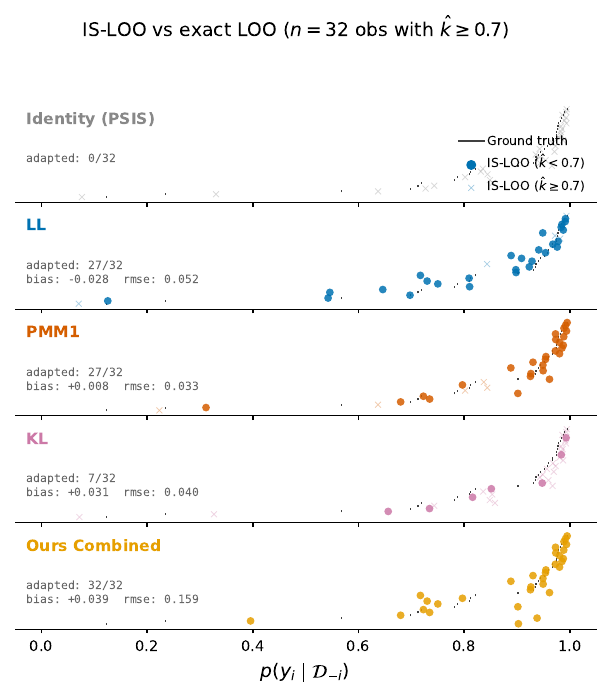}
	\caption{\textbf{Per-observation LOO-elpd for problematic observations} ($\hat{k}\geq 0.7$ under identity) in the ovarian logistic regression model. Each panel shows one IS-LOO method's estimate (colored) against the exact LOO ground truth (black). Observations are ordered by ground truth elpd. The IS-LOO methods closely track the ground truth, with the combined estimate (bottom panel) showing excellent agreement.}
	\label{fig:loo_validation}
\end{figure}

To validate the IS-LOO estimates produced by our transformations, we computed exact LOO cross-validation for the ovarian logistic regression model by refitting the Stan model 54 times (once per observation).
Fig.~\ref{fig:loo_validation} compares the per-observation LOO log predictive density ($\textrm{elpd}_i$) from each IS-LOO method against the ground truth, restricted to observations where adaptation was needed ($\hat{k} \geq 0.7$ under the identity transformation).
Table~\ref{tab:loo_validation} reports the total LOO-elpd and LOO-AUC for each method.

\begin{table}[h!]
	\centering
	\small
	\begin{tabular}{lccccc}
		\toprule
		\textbf{Method} & \textbf{Adapted} & \textbf{Total elpd} & \textbf{$\Delta$ vs GT} & \textbf{RMSE$_p$} & \textbf{LOO-AUC} \\
		\midrule
		Ground Truth (refit) & -- & $-14.06\pm3.01$ & -- & -- & $0.968$ \\
		Identity (PSIS) & $0/32$ & $-11.57\pm2.98$ & $+2.48\pm2.98$ & -- & $0.985$ \\
		LL & $27/32$ & $-14.19\pm3.41$ & $-0.13\pm3.41$ & $0.052$ & $0.958$ \\
		PMM1 & $27/32$ & $-12.07\pm3.00$ & $+1.98\pm3.00$ & $0.033$ & $0.981$ \\
		KL & $7/32$ & $-11.49\pm2.98$ & $+2.56\pm2.98$ & $0.040$ & $0.985$ \\
		PMM+LL+KL+Var & $31/32$ & $-13.50\pm3.40$ & $+0.55\pm3.40$ & $0.042$ & $0.971$ \\
		\bottomrule
	\end{tabular}
	\caption{\textbf{IS-LOO vs exact LOO} for the ovarian logistic regression model ($n=54$, $p=1536$, $s=1000$ posterior samples).
	``Adapted'' counts observations (of 32 with $\hat{k}\geq 0.7$) where the method achieves $\hat{k}<0.7$; for adapted observations, the method's estimate is used, otherwise the identity (PSIS) estimate is retained.
	Total elpd = $\sum_i \log p(y_i | \mathcal{D}_{-i})$; $\Delta$ = difference from ground truth; RMSE$_p$ = root mean squared error of the per-observation predictive probability $p(y_i|\calD_{-i})$ on successfully adapted observations only; LOO-AUC = area under the ROC curve using LOO predicted probabilities.
	LL achieves the closest total elpd to ground truth ($\Delta=-0.13$).
	The residual $+0.55$ elpd discrepancy in the combined row reflects PSIS bias on the 22 observations that did not require adaptation.}
	\label{tab:loo_validation}
\end{table}

\subsection{Ovarian cancer: Bayesian logistic regression}
\label{sec:suppresults_model}

\begin{figure}[h!]
	\centering
	\includegraphics[width=0.6\linewidth]{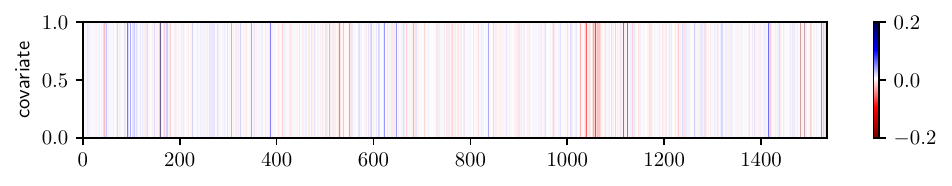}
	\caption{\textbf{Ovarian cancer logistic regression coefficients} for MCMC-based fitting.}
	\label{fig:lrmcmcfit}
\end{figure}

\begin{figure}[h!]
	\centering
	\includegraphics[width=0.65\textwidth]{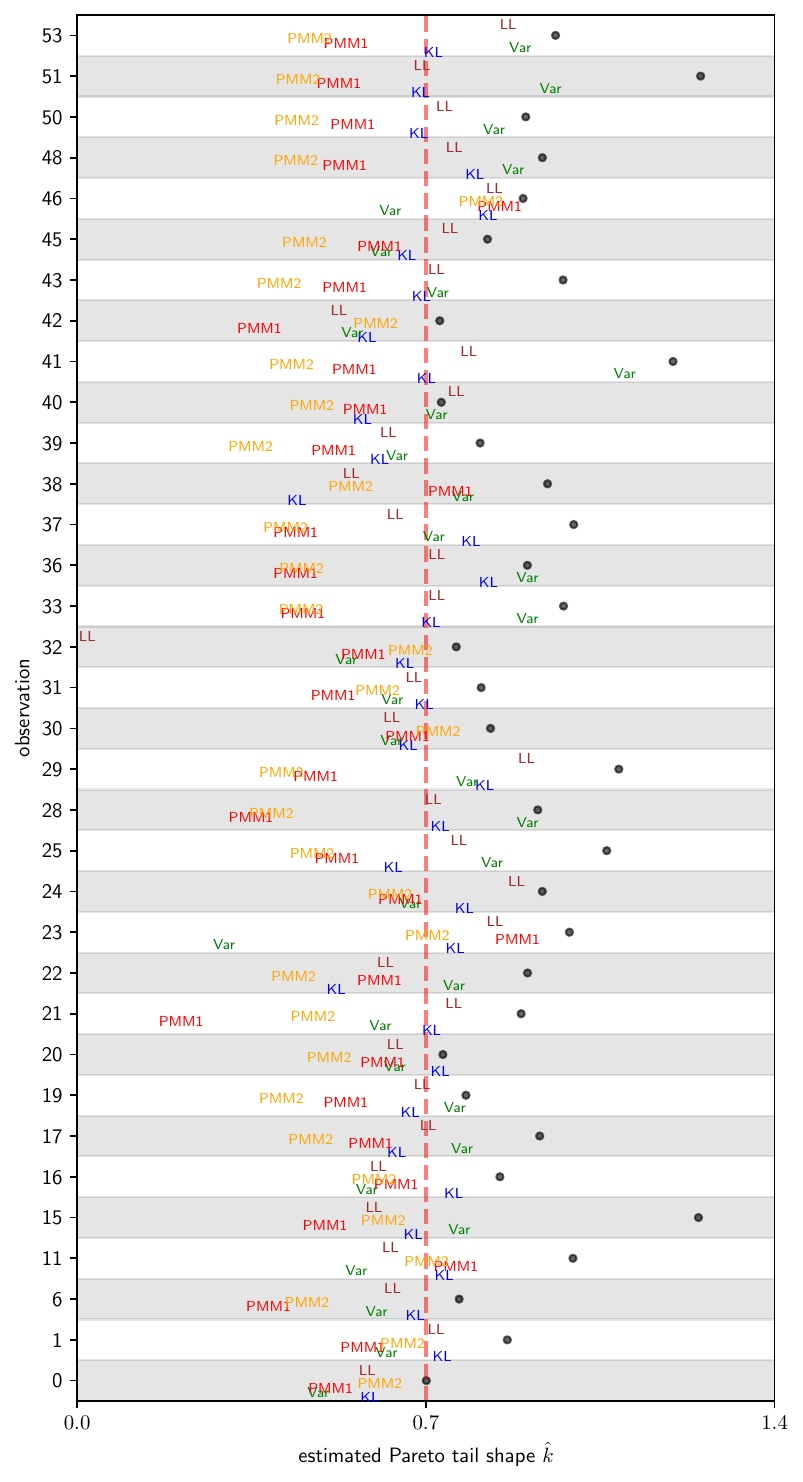}
	\caption{\textbf{Scatterplot of estimated Pareto tail shape diagnostic} \textbf{$\hat{k}$} versus observation, for transformed ovarian cancer logistic regression model parameters, for observations where the un-transformed samples have tail shape diagnostic $\hat{k}>0.7$ (black dots $\sbullet[0.95]).$ Values of minimum $\hat{k}$ for each transformation plotted:  blue for \textcolor{mplblue}{KL},  green for \textcolor{mplgreen}{Var}, red for \textcolor{mplred}{PMM1}, orange for \textcolor{mplorange}{PMM2}, purple for \textcolor{mplpurple}{MM1}, tan for \textcolor{mpltan}{MM2}, and brown for \textcolor{mplbrown}{LL} -- the minimum observed value for each transformation labeled. Adaptation for an observation is successful if $\hat{k}<0.7$ for any transformation.
		If the minimum value for a given transformation and observation falls outside of this displayed range then the corresponding point is omitted from this plot.
	}\label{fig:khat}
\end{figure}

\begin{figure}[h!]
	\centering
	\includegraphics[width=0.6\linewidth]{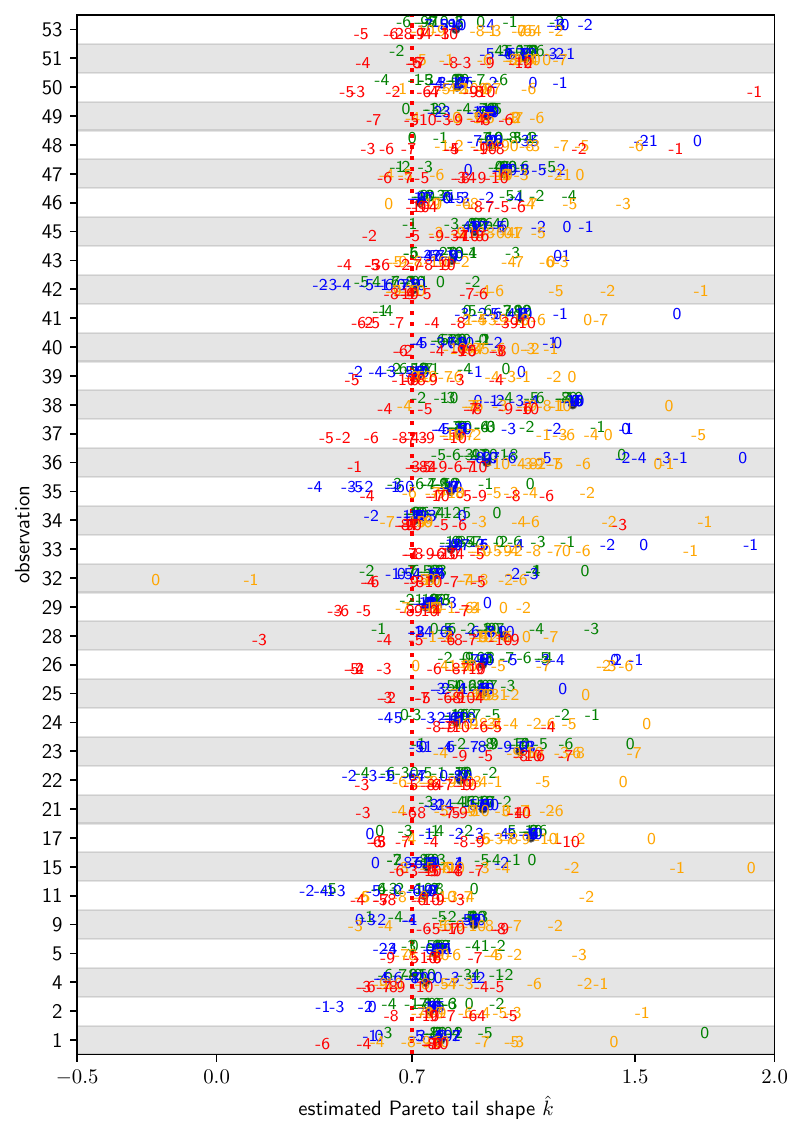}
	\caption{\textbf{Estimated Pareto tail shape diagnostic} \textbf{$\hat{k}$}, for logistic regression on the ovarian cancer dataset, for observations where  $\hat{k}>0.7$ (shown  as dots $\sbullet[0.75]).$ Post-transformed values of $\hat{k}$: blue for \textcolor{mplblue}{KL},  green for \textcolor{mplgreen}{Var}, red for \textcolor{mplred}{PMM1}, orange for \textcolor{mplorange}{PMM2}, purple for \textcolor{mplpurple}{MM1}, tan for \textcolor{mpltan}{MM2}, and brown for \textcolor{mplbrown}{LL} plotted, with location of the minimum observed value for each transformation labeled. Adaptation is successful if $\hat{k}<0.7.$ }\label{fig:khatdetailed}
\end{figure}

In Fig.~\ref{fig:lrmcmcfit} we display the posterior expectation for the regression coefficients.
Fig.~\ref{fig:khatdetailed} provides a more-detailed version of Fig.~\ref{fig:khat}, where each point is labeled by its corresponding value of $\log_4(\bar{h})$.

\section{Reproducibility: Jupyter notebooks}
\label{sec:notebooks}

The following pages contain Jupyter notebooks used for producing the results in this paper.
An implementation in Python/JAX is available at \texttt{github:mederrata/bayesianquilts}.

\subsection{Logistic regression}
\label{sec:jupyterlr}
\newpage
\includepdf[pages=-]{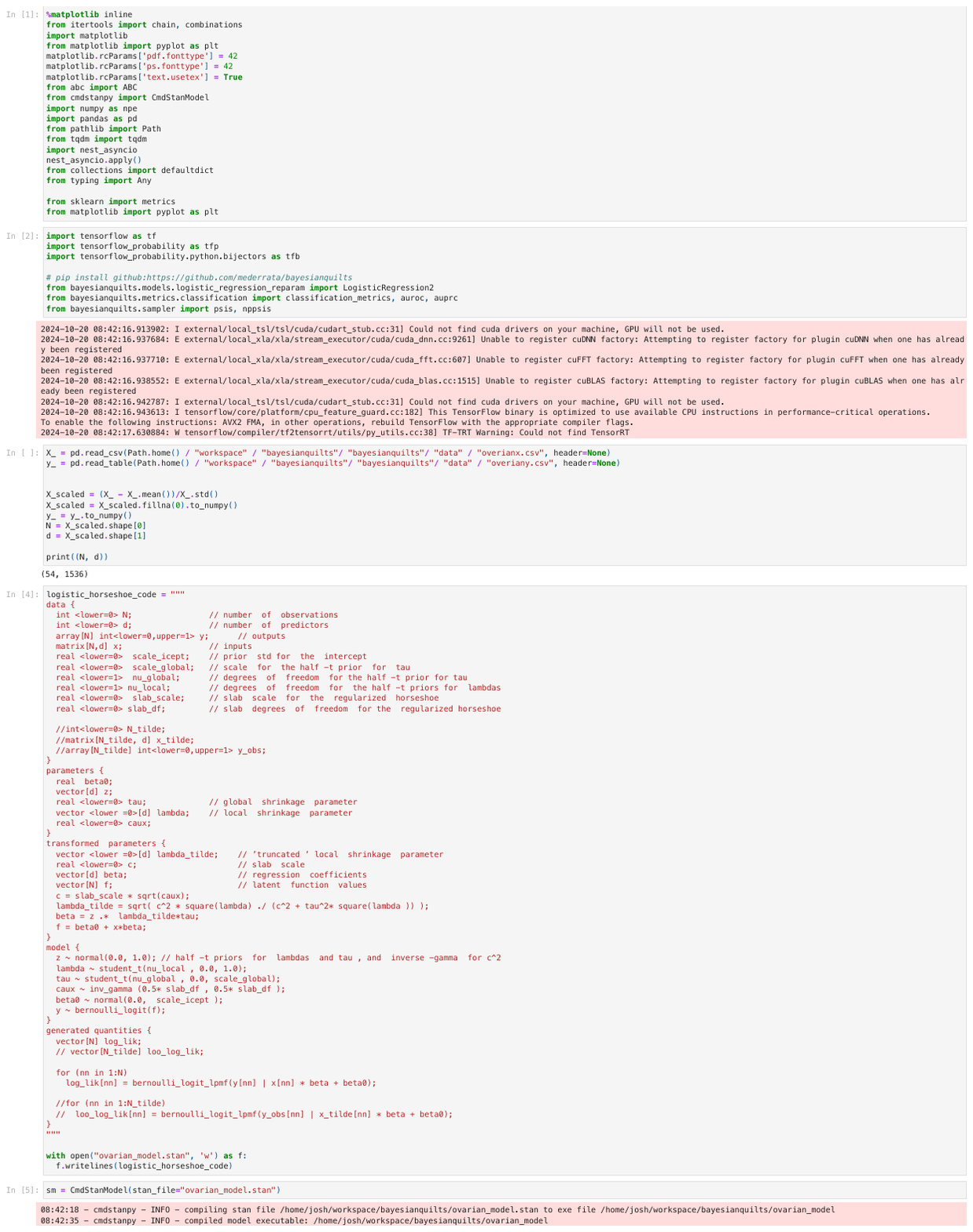}

\subsection{Shallow ReLU-net}
\label{sec:jupyterrelu}
\newpage
\includepdf[pages=-]{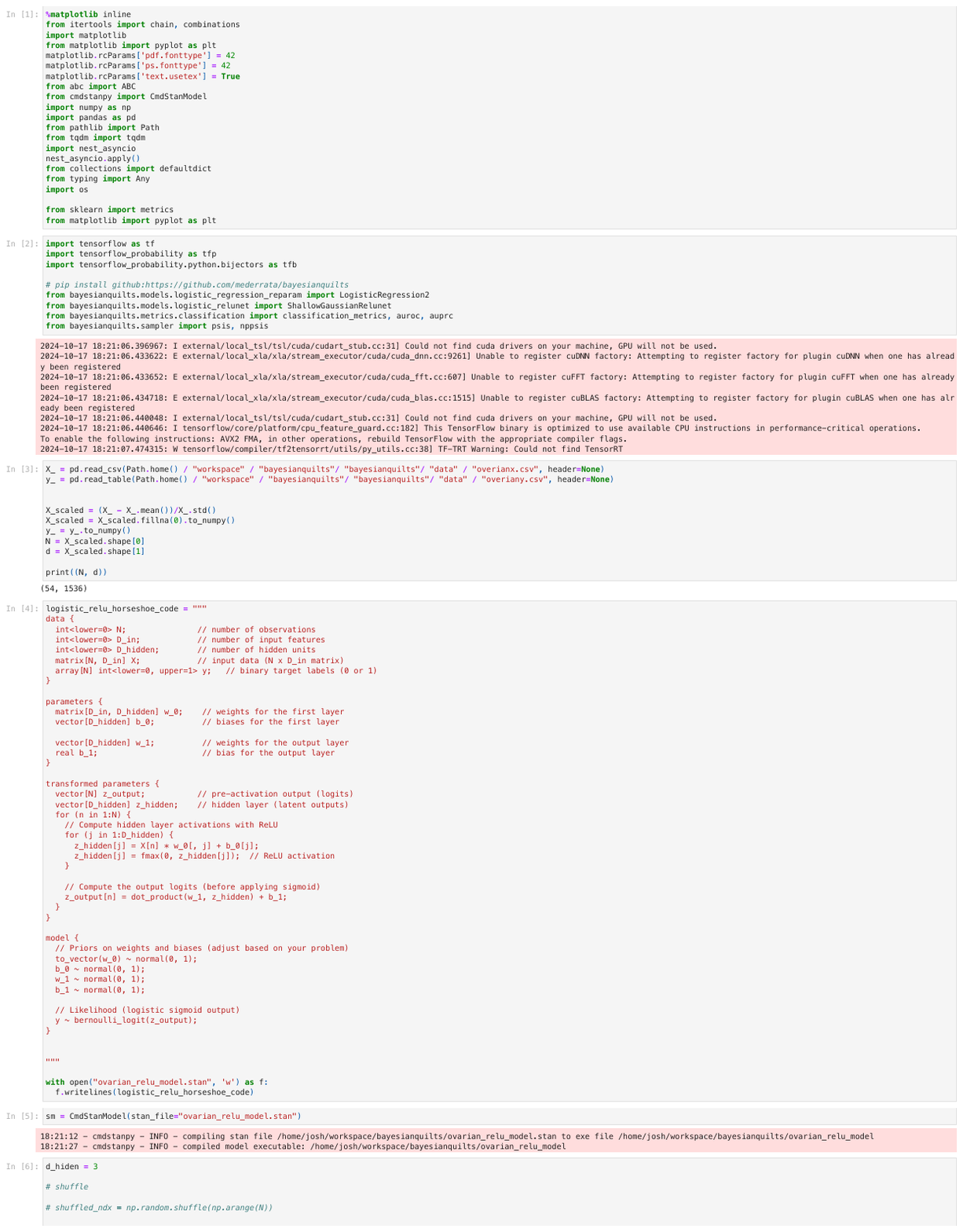}

\end{document}